\documentclass[a4paper,11pt]{article}
\pdfoutput=1 

\usepackage{jcappub} 

\usepackage[T1]{fontenc} 
\usepackage[utf8]{inputenc}

\usepackage{diagbox}
\usepackage{color}
\usepackage{multirow}
\usepackage{hhline}

\title{Gravitational-Wave Detector Networks: Standard Sirens on Cosmology and Modified Gravity Theory }

\author{Tao Yang}

\affiliation{Asia Pacific Center for Theoretical Physics, Pohang 37673, Korea}

\emailAdd{tao.yang@apctp.org}

\abstract{We construct the catalogues of standard sirens (StS) based on the future gravitational wave (GW) detector networks, i.e., the second-generation ground-based advanced LIGO+advanced Virgo+KAGRA+LIGO-India (HLVKI), the third-generation ground-based Einstein Telescope+two Cosmic Explorer (ET+2CE), and the space-based LISA+Taiji. From the corresponding electromagnetic (EM) counterpart detectors for each networks, we sample the joint GW+EM detections from the probability to construct the Hubble diagram of standard sirens for 10 years detections of HLVKI, 5 years detections of ET+2CE, and 5 years of detections of LISA+Taiji, which we estimate would be available and released in the 2030s. Thus we construct a combined Hubble diagram from these ground and spaced-based detector networks to explore the expansion history of our Universe from redshift 0 to 7. We give a conservative and realistic estimation of the catalogue and Hubble diagram of GW standard sirens and their potential on studying cosmology and modified gravity theory in the 2030s. We adopt two strategies for the forecasts. One is the traditional model-fitting Markov-Chain Monte-Carlo method (MCMC). The results show that the combined StS alone can constrain the Hubble constant at the precision level of $0.34\%$, 1.76 times more tightly than the current most precise measurement from \textit{Planck}+BAO+Pantheon. The joint StS with current EM experiments will improve the constraints of cosmological parameters significantly. The modified gravity theory can be constrained with $0.46\%$ error from the GW propagation. In the second strategy, we use the machine-learning nonparametric reconstruction techniques, i.e., the Gaussian process (GP) with the Artificial Neural Networks (ANN) as a comparison. GP reconstructions can give comparable results with MCMC. We anticipate more works and research on these topics.}


\begin{document}
\maketitle
\flushbottom


\section{Introduction}

The first observation of gravitational wave (GW) from a binary black holes merger~\cite{Abbott:2016blz} and the following series of detections~\cite{LIGOScientific:2018mvr,Abbott:2020niy}, especially the first joint observation of GW from a binary neutron stars (BNS) with its electromagnetic counterpart (EM)~\cite{TheLIGOScientific:2017qsa,GBM:2017lvd,Monitor:2017mdv}, have opened the new era of multimessenger astronomy. Gravitational waves, as the novel signals in our Universe compared to the traditional EM observations, have sparked a series of research on cosmology, astrophysics and fundamental physics (see e.g.~\cite{TheLIGOScientific:2016src,TheLIGOScientific:2016htt,Abbott:2017xzu,LIGOScientific:2019fpa,Abbott:2020jks,Hotokezaka:2018dfi,Ezquiaga:2017ekz,Baker:2017hug,Copeland:2018yuh,Abbott:2021ksc}). GWs play significant roles in studying such as the physics of early Universe, the expansion history of the late-time Universe, the nature of dark energy and dark matter, the properties of black holes, and the test of general relativity (GR) (see reviews e.g.~\cite{Schutz:1999xj,Barack:2018yly,Sasaki:2018dmp,Gair:2012nm,Ezquiaga:2018btd,Cai:2017cbj,Meszaros:2019xej,Christensen:2018iqi,Perkins:2020tra}). With advanced LIGO and advanced Virgo reaching their target sensitivity, and other detectors such as KAGRA and LIGO-India joining the search in the near future, the second-generation (2G) ground-based detector network HLVKI (consists of advanced LIGO-Hanford, advanced LIGO-Livingston, advanced Virgo, KAGRA and LIGO-India) would provide remarkable measurements of Hubble constant within a few years~\cite{Chen:2017rfc}. On a longer timescale, around the 2030s the third-generation ground-based detectors such as the Einstein Telescope (ET)~\footnote{\url{http://www.et-gw.eu/}} in Europe and Cosmic Explorer (CE)~\footnote{\url{https://cosmicexplorer.org/}} in the US, and the space interferometer LISA~\footnote{\url{https://www.lisamission.org/}} will have the potential of detecting a large number of coalescing compact binaries at cosmological redshifts. During the same period, Chinese space-based GW detector which is proposed by the Chinese Academy of Sciences (CAS) as ``Taiji Program in Space'' would be launched~\cite{Hu:2017mde}. Thus we expect the construction of  GW detector networks from ground to space would be reality in the 2030s. These GW detector networks' synergetic potential on cosmology, astrophysics and fundamental physics deserve detailed research and investigations. 

The traditional EM observations such as cosmic microwave background (CMB), type Ia supernovae (SNe Ia), 
baryon acoustic oscillations (BAO) and the large scale structures have already depicted our Universe with great information~\cite{Riess:1998cb,Perlmutter:1998np,Hinshaw:2012aka,Aghanim:2018eyx,Alam:2016hwk}. The standard cosmology with the late-time Universe formulated by the so called $\Lambda$CDM model is favored by most experiments despite several discordances such as the Hubble tensions     
(see reviews of~\cite{Freedman:2017yms,Verde:2019ivm} and recent review of the solutions~\cite{divalentino2021realm}) and cosmic shear discrepancies~\cite{McCarthy:2017csu,Hildebrandt:2018yau,Asgari:2019fkq}. The former is crucial to understanding the current expansion rate of our Universe and it challenges the standard cosmological models and even the fundamental physics~\cite{Riess:2019cxk}. A third-party measurement of the Hubble constant independent of high redshift CMB and local SNe Ia is thus significantly important. Gravitational wave can provide measurements of $H_0$ in the medium redshift range. It is among the most promising tools to resolve this puzzle. Using GWs to measure the Hubble constant was proposed by~\cite{Schutz:1986gp} based on the fact that the luminosity distance can be directly inferred from the GW waveform. Analogously to the SNe Ia ``standard candles'' one calls this feature of GWs as ``standard sirens'' (StS), while the latter is independent of calibration but limited by the redshift measurement of the source. In this paper we focus on the ``bright sirens'', i.e., the redshift is measured with the help of the EM counterparts. The first standard sirens measurement of Hubble constant is from GW170817 which is emitted from a BNS system and accompanied by a gamma-ray burst (GRB 170817A)~\cite{Abbott:2017xzu}. Though current measurement is not precise enough to resolve the Hubble tension, the standard sirens from future HLVKI network are expected to contribute more on this issue~\cite{Hu:2017mde}. 

Beside the Hubble tension, GW standard sirens also aim at studying the nature of dark energy and modified gravity (MG) theory. The 3G ground-based detector ET can detect the BNS or BH-NS binaries up to redshift 2--3 thus is very helpful for constraining the dynamics of dark energy~\cite{Sathyaprakash:2009xt,Zhao:2010sz,Cai:2016sby}, as well as MG~\cite{Belgacem:2019tbw}.  For LISA, the higher redshift ($z\sim 6-7$) standard sirens of the massive black hole binaries (MBHBs) can be approached and several detailed investigations have been made on these topics~\cite{Tamanini:2016zlh,Tamanini:2016uin,Caprini:2016qxs,Belgacem:2019pkk}. In addition to HLVKI, the 3G ground-based network ET+2CE~\footnote{\url{https://gwic.ligo.org/}} has also been studied in several literature~\cite{Belgacem:2019tbw,Sathyaprakash:2019rom}. Recently, the network of space-based detectors, i.e., LISA-Taiji has been proposed and several perspectives such as the improvement of the precision of the source localization~\cite{Ruan:2019tje,Ruan:2020smc,Wang:2020vkg} and the applications on cosmology have been investigated (see e.g.~\cite{Wang:2020dkc,Orlando:2020oko,Wang:2021srv}). These works showed the potential of the joint LISA+Taiji network on studying the Universe.

\begin{figure}
\centering
\includegraphics[width=0.9\textwidth]{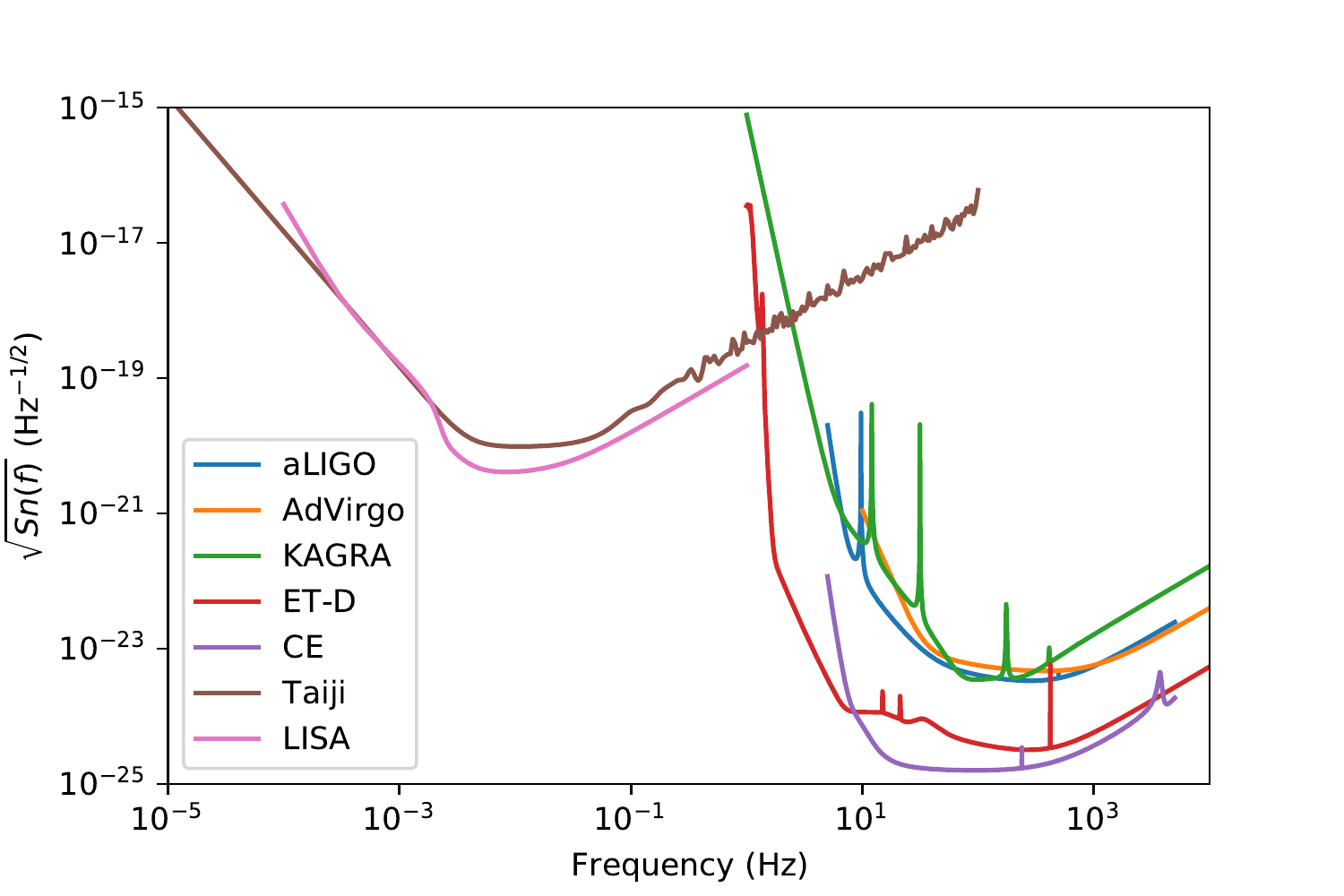}
\caption{The noise power spectral density of the GW detector networks.}
\label{fig:psd}
\end{figure}

In this paper, based on previous research especially~\cite{Tamanini:2016zlh,Belgacem:2019tbw}, we would like to construct the catalogues and Hubble diagrams of GW standard sirens by future ground/space-based GW detector networks, i.e., HLVKI; ET+2CE; LISA+Taiji. Figure~\ref{fig:psd} shows the sensitivity curves of these GW detectors and we can see different networks aim at varying frequency bands of GW and sources.  Previous research showed that the redshifts of standard sirens for 2G/3G ground-based and space-based detectors are distributed separately but continually. Thus intuitively one could combine them together to build the cosmic distance ladders from the local Universe to cosmic distance around $z\sim 7$. Considering the fact that ET, CE, LISA and Taiji would be launched during the 2030s, we would like to give a realistic construction of the catalogue and Hubble diagram of standard sirens which is expected to be a reality in the 2030s.  The applications of the Hubble diagram on cosmology and modified gravity theory will be investigated. 

The structure of this paper is as follows. In section~\ref{sec:StScons} we show the details of constructing the Hubble diagram of standard sirens by 2G ground-based networks HLVKI and 3G ET+2CE, and space-based network LISA+Taiji. For every network we construct the catalogue of the GW events by the threshold of detection and also calculate the probability of the joint GW+EM events according to the limits of the EM counterpart detectors. Then from the probability we construct the standard sirens catalogues for each networks and finally combine them together to draw a whole Hubble diagram. In section~\ref{sec:StSCosMG} we apply the combined Hubble diagram to cosmology and test of GR. We consider two different strategies. The first is the traditional model-fitting method by which we can explore the space of  cosmological parameters using Markov-Chain Monte-Carlo (MCMC). Combining  different data sets including CMB, BAO and SNe Ia, we constrain the cosmological parameters in the base $\Lambda$CDM model and its extensions like the dynamic dark energy models and also a phenomenological parametrization of MG through GW propagation. In the second strategy, we adopt the fashionable nonparametric reconstruction techniques which belong to machine learning to reconstruct the luminosity distance of standard sirens as a function of redshift. We use two different state-of-the-art techniques. One is the widely-used Gaussian process (GP) from which we reconstruct the Hubble parameter and equation of state of dark energy. The modified propagation of GW is also analyzed by this nonparametric approach. The other one is the fast-developing Artificial Neural Networks (ANN) from which we  reconstruct the luminosity distance as a comparison with that of GP.
We then summarize our results and finally give our discussions and prospects in section~\ref{sec:dis}.


\section{Construction of catalogues of standard sirens from future GW detector networks \label{sec:StScons}}

The standard sirens we consider in this paper are the so called ``bright sirens'' which are GWs accompanied with EM counterparts. For 2G/3G ground-based GW detectors we focus on the BNS mergers, from which the EM counterparts like GRBs are expected to be observed. Note we do not include the NS-BH binaries since they are only a small fraction of the BNS from the merger rates predicted in~\cite{LIGOScientific:2018mvr}. Several research has argued that the NS-BH binaries can break the degeneracy between the luminosity distance and inclination angle due to the BH spin precession and hence significantly improve the estimation of luminosity distance compared to BNS system~\cite{Vitale:2018wlg,Vitale:2014mka,Feeney:2020kxk}. However, considering the expected number of such events, in this paper we do not take them into account when constructing the whole Hubble digram. In the millihertz frequency range, the space-based GW detector LISA is going to detect MBHB mergers at cosmological distances (up to redshift 15--20). MBHBs are expected to produce powerful EM counterparts, since they are believed to merge in a gas rich environment that may power EM emission through jets, disk winds or accretion: this fact will allow us to determine precisely the object positions in the sky.
Several studies found that MBHBs could emit radiation in different bands of the EM spectrum at inspiral, merger and during long lasting (ranging from weeks to months) afterglows~\cite{Kocsis:2007yu,OShaughnessy:2011nwl,Kaplan:2011mz,Palenzuela:2010nf,Dotti:2011um,Giacomazzo:2012iv,Haiman:2017szj}. The EM counterparts could be observed up to $z\sim7$~\cite{Tamanini:2016zlh}. For the space-based LISA+Taiji network we focus on the MBHB standard sirens. 
In this paper we would like to construct the catalogues of standard sirens from future GW detector networks in the timing of the 2030s. Considering the launch time of these networks we choose the operation time to be 10 years for HLVKI, 5 years for ET+2CE and 5 years for LISA+Taiji. 


\subsection{BNS standard sirens based on ground-based 2G HLVKI and 3G ET+2CE}

To construct the mock catalogues of standard sirens from the inspiral of BNS by future ground-based 2G HLVKI and 3G ET+2CE networks we first need the merger rate of BNS.  We follow~\cite{Vitale:2018yhm,Belgacem:2019tbw} (and reference therein) for the star formation rate (SFR) and the time delay between the formation of BNS progenitors and their eventual merger. The merger rate density per unit redshift in the observer frame can be expressed as
\begin{equation}
R_z(z)=\frac{R_m(z)}{1+z}\frac{dV(z)}{dz} \,,
\label{eq:Rz}
\end{equation}
where $dV/dz$ is the comoving volume element and $R_m$ is the rate per volume in the source frame. The merge rate per volume at redshift $z_m$ is related to the formation rate of massive binaries through the time delay distribution $P(t_d)$,
\begin{equation}
R_m(z_m)=\int_{z_m}^{\infty}dz_f\frac{dt_f}{dz_f}R_f(z_f)P(t_d) \,.
\label{eq:Rm}
\end{equation}
All the systems that merge at a look-back time $t_m$ (or redshift $z_m$) are systems that formed at $t_f$ (or redshift $z_f$) and $t_d=t_f-t_m$ is the time delay. $R_f$ is the formation rate of massive binaries and we assume it is proportional to the Madau-Dickinson (MD) star formation rate~\cite{Madau:2014bja},
\begin{equation}
\psi_{\rm MD}=\psi_0\frac{(1+z)^{\alpha}}{1+[(1+z)/C]^{\beta}} \,,
\label{eq:psiMD}
\end{equation}
with parameters $\alpha=2.7$, $\beta=5.6$ and $C=2.9$. The coefficient $\psi_0$ is the normalization factor which is determined by the BNS rate we set at $z=0$. Here we adopt $R_m(z=0)=920~\rm{Gpc}^{-3}\rm{yr}^{-1}$ which is the median rates estimated from the O1 LIGO observation run and the O2 LIGO/Virgo observation run~\cite{LIGOScientific:2018mvr} and assume a Gaussian distribution of the mass of neutron stars (for a flat mass distribution one gets a slightly lower $R_m(z=0)=662~\rm{Gpc}^{-3}\rm{yr}^{-1}$). This is also consistent with the latest first half O3 run~\cite{Abbott:2020niy}. For the time delay distribution $P(t_d)$ we follow~\cite{Vitale:2018yhm} and adopt the exponential form with an e-fold time of $\tau=100$ Myr,
\begin{equation}
P(t_d,\tau)=\frac{1}{\tau}\exp(-t_d/\tau) \,.
\end{equation}

\begin{figure}
\centering
\includegraphics[width=0.9\textwidth]{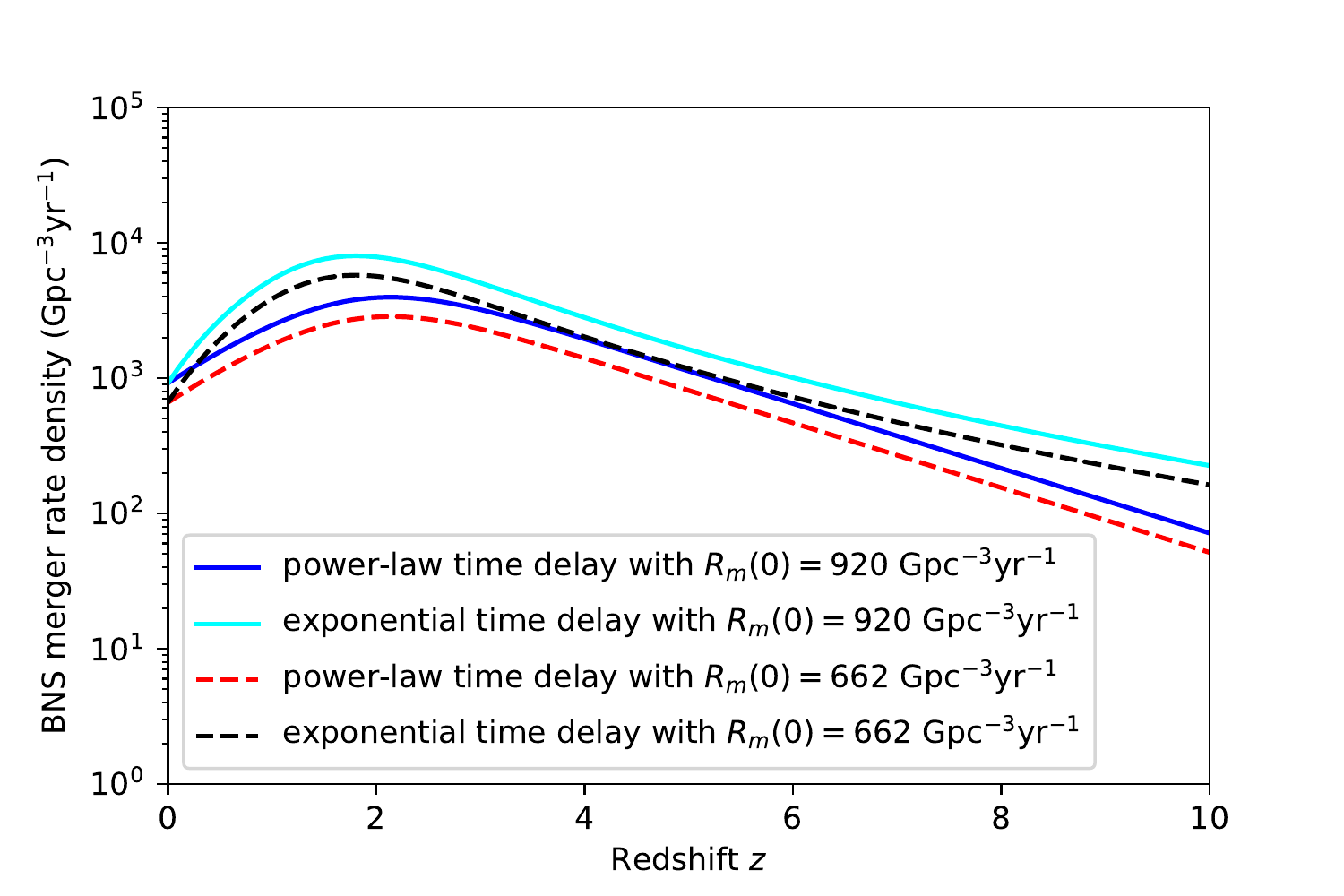}
\caption{The BNS merger rate density along the redshift with different models of the star formation rate, time delay distribution and local BNS merger rate. In this paper we use the exponential time delay distribution with the local BNS merger rate $920~\rm{Gpc}^{-3}\rm{yr}^{-1}$. }
\label{fig:Rm}
\end{figure}

Some different choices of such as SRF function, the distribution function of time-delay, and the local BNS rates are adopted in~\cite{Vitale:2018yhm,Belgacem:2019tbw}. For example~\cite{Belgacem:2019tbw} uses the SFR of~\cite{Vangioni:2014axa} and assumes the  power-low form of the distribution of time delay $P(t_d)\sim t_d^{\alpha}$ with $\alpha=-1$ and a minimum delay time 20 Myr for a massive binary to evolve until coalescence. Figure~\ref{fig:Rm} shows the normalized BNS merger rate density for four different scenarios. Actually the slight difference would not influence our final results~\footnote{The differences between different models are within 1 order of magnitude. As we can see later, the number of standard sirens is actually limited by the EM observations and only a very small fraction of the GW events can be finally identified to be the standard sirens.}.

Having the BNS merger rate we can easily draw the distribution of the redshift of the GW sources. To select the GW events produced from the BNS system based on a given GW detector network one needs to calculate the signal-to-noise ratio (SNR). From the inspire part of the signal, the SNR detected by matched filtering with an optimum filter in the ideal case of Gaussian noise is given by
\begin{equation}
\rho^2=\frac{5}{6}\frac{(G\mathcal{M}_c)^{5/3}\mathcal{F}^2}{c^3\pi^{4/3}d_L^2(z)}\int^{f_{\rm max}}_{f_{\rm min}}df\frac{f^{-7/3}}{S_n(f)} \,,
\label{eq:SNR}
\end{equation}
where $\mathcal{M}_c$ is the redshifted chirp mass which is a combination of two component masses, $\mathcal{M}_c=(m_1m_2)^{3/5}/(m_1+m_2)^{1/5}(1+z)$. $d_L$ is the luminosity distance. $S_n(f)$ is the one-sided noise power spectral density (PSD) of detector. The factor $\mathcal{F}$ is to characterize the detector response, $\mathcal{F}^2=\frac{(1+\cos^2\iota)^2}{4}F^2_++\cos^2\iota F^2_\times$. $F_+$ and $F_\times$ are the antenna response functions to the GW + and $\times$ polarizations. For the ``L-shape'' detectors like LIGO, Virgo, KAGRA, and CE the antenna response functions are
\begin{align}
&F_+=\frac{1}{2}(1+\cos^2\theta)\cos 2\phi \cos 2\psi-\cos \theta \sin 2\phi \sin 2\psi  \,, \\ 
&F_\times=\frac{1}{2}(1+\cos^2\theta)\cos 2\phi \sin 2\psi+\cos \theta \sin 2\phi \cos 2\psi \,.
\end{align}
For the triangle-shape ET the antenna response functions include a extra factor $\sqrt{3}/2$ compared to LIGO. Since ET has three independent detectors, the other two pairs of $F_+$ and $F_\times$ are just the same as the first one except $\phi$ is replaced by $\phi+2\pi/3$ and $\phi+4\pi/3$. For the range of the frequency, we choose $f_{\rm min}$ to be the low frequency limit of the detector and $f_{\rm max}$ to be the redshifted frequency of the Inner-most Stable Circular Orbit (ISCO), i.e., $f_{\rm max}=2f_{\rm ISCO}=\frac{1}{3\sqrt{6}(2\pi)}\frac{c^3}{G(m_1+m_2)(1+z)}$, at which the inspiral phase ends~\cite{Maggiore:GW1}.

\begin{figure}
\centering
\includegraphics[width=0.9\textwidth]{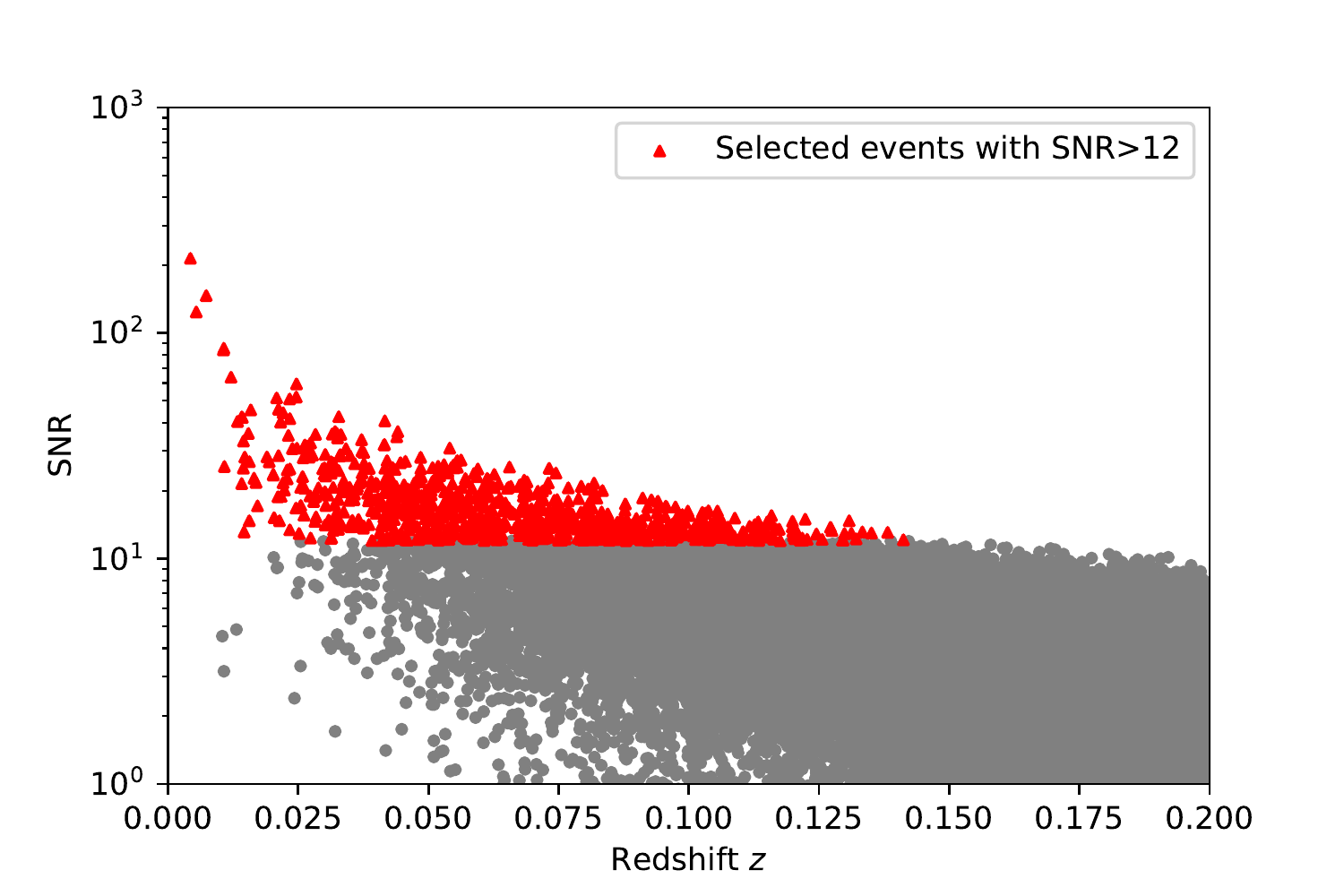}
\caption{A realization of the mock catalogue of the 10 yr BNS GW detections from HLVKI network. Here we show the scatter plots of SNR against the redshift.}
\label{fig:HLVKI_scatter}
\end{figure}

For every redshift of source drawn from the distribution of BNS merger rate, we assign the sky location ($\theta$, $\phi$) and inclination angle $\iota$ from isotropic distribution.  The polarization $\psi$, component masses of BNS are drawn from the uniform distribution, i.e., $\psi\in[0,2\pi)$ and $m_1,m_2\in[1,2]M_{\odot}$. Having sampled all the parameters one can calculate the SNR of each GW candidates from~\eqref{eq:SNR}. The coherent SNR, assuming uncorrelated noises among the detectors, is simply given by the quadrature sum of the individual SNRs, $\rho^2_{\rm tot}=\Sigma_i\rho^2_i$. We assume that each detector has a duty cycle of $80\%$. In addition, we draw the ``measured'' SNR from a Gaussian distribution centered at the matched-filter value with a standard deviation $\sigma=1$~\cite{Chen:2017rfc}. We then classify the event as detectable if the combined SNR among the detectors in the network is larger than a threshold $\rho_t=12$~\cite{Chen:2017rfc,Belgacem:2019tbw}.
Figure~\ref{fig:HLVKI_scatter} shows an example of the 10 yr detections of BNS from HLVKI network. We find that the number of BNS mergers detected by HLVKI network is around 86/yr which is very close to the number estimated by~\cite{Belgacem:2019tbw} and also in the same order of that given by~\cite{Sathyaprakash:2019rom}.

To mock up the standard sirens from the BNS mergers we first need the luminosity distance measurements inferred from the matched filtering waveform. We assume the fiducial cosmology to be the $\Lambda$CDM model with $H_0=67.72~\rm km~s^{-1}~Mpc^{-1}$ and $\Omega_m=0.3104$, corresponding to the mean values obtained from the latest \textit{Planck} TT,TE,EE+lowE+lensing+BAO+Pantheon data combination~\cite{Aghanim:2018eyx}. We also fix the present CMB temperature $T_{\rm CMB}=2.7255~\rm K$, the sum of neutrino masses $\Sigma_{\nu}m_{\nu}=0.06~\rm eV$, and the effective extra relativistic degrees of freedom $N_{\rm eff}=3.046$, as in the {\it Planck} baseline analysis.   The simulated luminosity distance is sampled from the Gaussian distribution $\mathcal{N}(d_L^{\rm fid}, \Delta d_L)$. The center value $d_L^{\rm fid}$ is calculated from the fiducial cosmology. The error $\Delta d_L$ includes the instrumental error and some other errors from such as weak lensing (for some recent discussions of lensing effects on GW see e.g.~\cite{Bertacca:2017vod,Cusin:2020ezb,Cusin:2019eyv,Shan:2020esq}) and peculiar velocity of the source galaxy. For the weak lensing we adopt the analytically fitting formula~\cite{Hirata:2010ba,Tamanini:2016zlh}
\begin{equation}
\left(\frac{\Delta d_L(z)}{d_L(z)}\right)_{\rm lens}=0.066\left(\frac{1-(1+z)^{-0.25}}{0.25}\right)^{1.8} \,.
\end{equation}
Note for ET+2CE and LISA+TAIJI networks that will work around the 2030s we consider a delensing factor, i.e., the use of dedicated matter surveys along the line of sight of the GW event in order to estimate the lensing magnification distribution and thus remove part of the uncertainty due to weak lensing, which can reduce the weak lensing uncertainty. Following~\cite{Speri:2020hwc} we adopt a phenomenological formula
\begin{equation}
F_{\rm delens}(z)=1-\frac{0.3}{\pi/2}\arctan(z/z_*) \,,
\end{equation}
where $z_*=0.073$. The final lensing uncertainty on $d_L$ will thus be
\begin{equation}
\left(\frac{\Delta d_L(z)}{d_L(z)}\right)_{\rm delens}=F_{\rm delens}(z)\left(\frac{\Delta d_L(z)}{d_L(z)}\right)_{\rm lens} \,.
\end{equation}
For the peculiar velocity uncertainty, we use the fitting formula~\cite{Kocsis:2005vv},
\begin{equation}
\left(\frac{\Delta d_L(z)}{d_L(z)}\right)_{\rm pec}=\left[1+\frac{c(1+z)^2}{H(z)d_L(z)}\right]\frac{\sqrt{\langle v^2\rangle}}{c} \,,
\end{equation}
here we set peculiar velocity value to be 500 km/s, in agreement with average values observed in galaxy catalogs.
Finally the instrumental error due to the parameter estimation from the matched filtering waveform, is estimated as 1/SNR~\cite{Dalal:2006qt,Li:2013lza}. Since in this paper we consider the BNS with the short GRB as the standard sirens, the short GRB is usually beamed within an angle of about $25^\circ$.  One can find that the correlation between distance and inclination is substantially broken, and the above estimate becomes accurate~\cite{Nissanke:2009kt}. Now the final uncertainty of the luminosity distance is just the sum of the above errors in quadrature.

To use the BNS as the standard sirens one also needs the redshift information of GW sources. In this paper we focus on the electromagnetic counterpart of GW to infer the redshift information. The identification of the EM counterpart is either from the follow-up EM observation based on an accurate localization from GW detector network (like GW170817) or just a temporal coincidence of the GW event with a short GRB. From many short GRBs, the redshift has indeed been determined from the X-ray afterglow, that can be accurately localized by \textit{Chandra} or \textit{Swift}/XRT. Following~\cite{Belgacem:2019tbw} and based on the working time of GW detector networks, we adopt different GRB satellites and telescopes to mock up the detections of joint GW+GRB for 2G HLVKI and 3G ET+2CE networks separately.

For a GRB detected in coincidence with a GW signal we require that the peak flux is above the flux limit of the satellite. Based on the results of~\cite{Howell:2018nhu} for fitting GRB170817A, we assume the Gaussian structured jet profile model
\begin{equation}
L(\theta_{\rm V})=L_c\exp(-\frac{\theta_{\rm V}^2}{2\theta_c^2}) \,,
\label{eq:Ltheta}
\end{equation}
with $L(\theta)$ the luminosity per unit solid angle, $\theta_{\rm V}$ the viewing angle and $L_c$ and $\theta_c$ the structure parameters that define the sharpness of the angular profile. The structured jet parameter is given by $\theta_c=4.7^\circ$. We then assume a standard broken power law of the form for the distribution of the short GRB
\begin{equation}
\Phi(L)\propto
\begin{cases}
(L/L_*)^a, & L<L_* \\
(L/L_*)^b, & L\ge L_*
\end{cases}
\label{eq:PhiL}
\end{equation}
where $L$ is the isotropic rest frame luminosity in the 1-10000 keV energy range and $L_*$ is a characteristic luminosity that separates the low and high end of the luminosity function and $a$ and $b$ are the characteristic slopes describing these regimes, respectively.  Following~\cite{Wanderman:2014eza} we have $a=-1.95$, $b=-3$ and $L_*=2\times 10^{52}~\rm erg~sec^{-1}$. We also assume a standard low end cutoff in luminosity of $L_{\rm min}=10^{49}~\rm erg~sec^{-1}$.
From the GW catalogue which has passed the threshold $\rho_{\rm tot}>12$ we can select the GW-GRB coincidences according to the probability distribution $\Phi(L)dL$. To calculate the probability of the GRB detection for every GW event we need to convert the flux limit of the GRB satellite $P_{\rm T}$ to the peak luminosity $L$ in~\eqref{eq:PhiL}. 

We begin from investigating the probability of detecting a GW signal by the HLVKI network in coincidence with a GRB by the current generation of GRB satellites. We assume here that the Fermi-GBM can make a coincident detection and that \textit{Swift} can slew to the combined GW/GRB error box and identify an X-ray counterpart. A Fermi-GBM detection is recorded if the value of the observed flux is greater than the flux limit $P_{\rm T}=1.1~\rm ph~sec^{-1}~cm^{-2}$  in the 50-300 keV band for Fermi-GBM~\cite{Howell:2018nhu}. We consider the standard flux-luminosity relation with two corrections: an energy normalization and a k-correction~\cite{Wanderman:2014eza,Howell:2018nhu}
\begin{equation}
b=\frac{\int^{1}_{10000}EN(E)dE}{\int^{E2}_{E1}N(E)dE} \,,
\end{equation}
\begin{equation}
k=\frac{\int^{E2}_{E1}N(E)dE}{\int^{E2(1+z)}_{E1(1+z)}N(E)dE} \,,
\end{equation}
where $[E1,E2]$ is the detector's energy window. The observed photon flux is scaled by $b$ to account for the missing fraction of the gamma-ray energy seen in the detector band. The cosmological k-correction is due to the redshifted photon energy when traveling from source to detector. $N(E)$ is the observed GRB photon spectrum.
For short GRBs we model the function $N(E)$ by the Band function~\cite{Band:2002te} which is a phenomenological fit to the observed spectra of GRB prompt emissions and is a function of spectral indices $(\alpha_B, \beta_B)$ and break energy, $E_b$, where the two power laws combine
\begin{equation}
N(E)=\begin{cases}
N_0\left(\frac{E}{100~\rm keV}\right)^{\alpha_B}\exp(-\frac{E}{E_0}), &~ E\leq E_b \\
N_0\left(\frac{E_b}{100~\rm keV}\right)^{\alpha_B-\beta_B}\exp(\beta_B-\alpha_B) \left(\frac{E}{100~\rm keV}\right)^{\beta_B}, &~E>E_b
\end{cases}
\end{equation}
here $E_b=(\alpha_B-\beta_B)E_0$ and $E_p=(\alpha_B+2)E_0$. From~\cite{Wanderman:2014eza}, we take $\alpha_B=-0.5$, $\beta_B=-2.25$ and a peak energy $E_p=800~\rm keV$ in the source frame. 
Following the relation between flux and luminosity for GRB~\cite{Meszaros:1995dj,Meszaros:2011zr}, one can convert the flux limit $P_T$ to the observed peak luminosity by
\begin{equation}
L=4\pi d_L^2(z)k(z)b/(1+z)P_T \,.
\end{equation}
Then from~\eqref{eq:Ltheta} and~\eqref{eq:PhiL} one can calculate the probability of each selected GW event which can have a GRB coincidence. Note we further assume the total time-averaged observable sky fraction of the Fermi-GBM, which is 0.60~\cite{Burns:2015fol}. Finally from the probability we can sample the GW-GRB detections from the total GW events.

\begin{figure}
\centering
\includegraphics[width=0.49\textwidth]{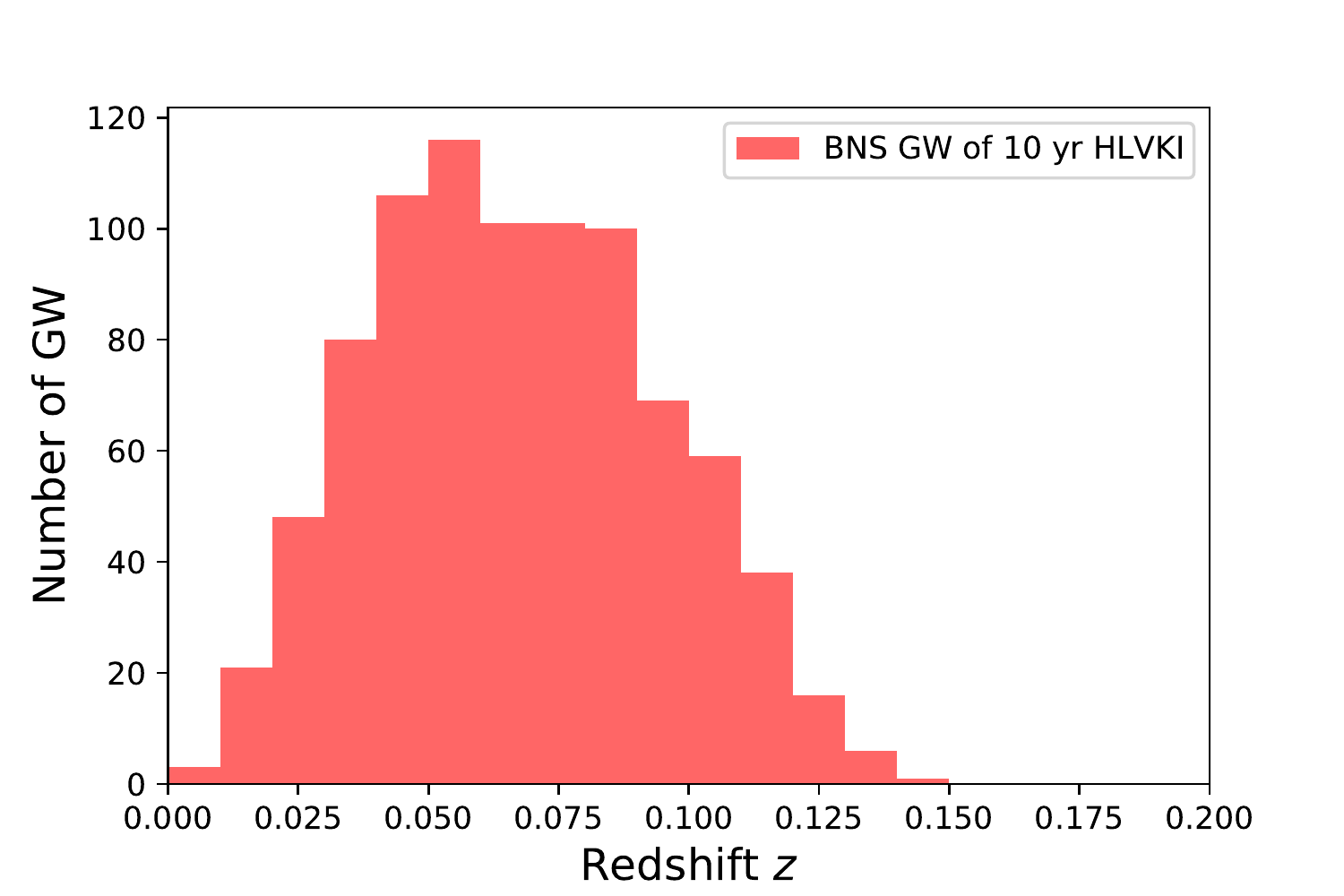}
\includegraphics[width=0.49\textwidth]{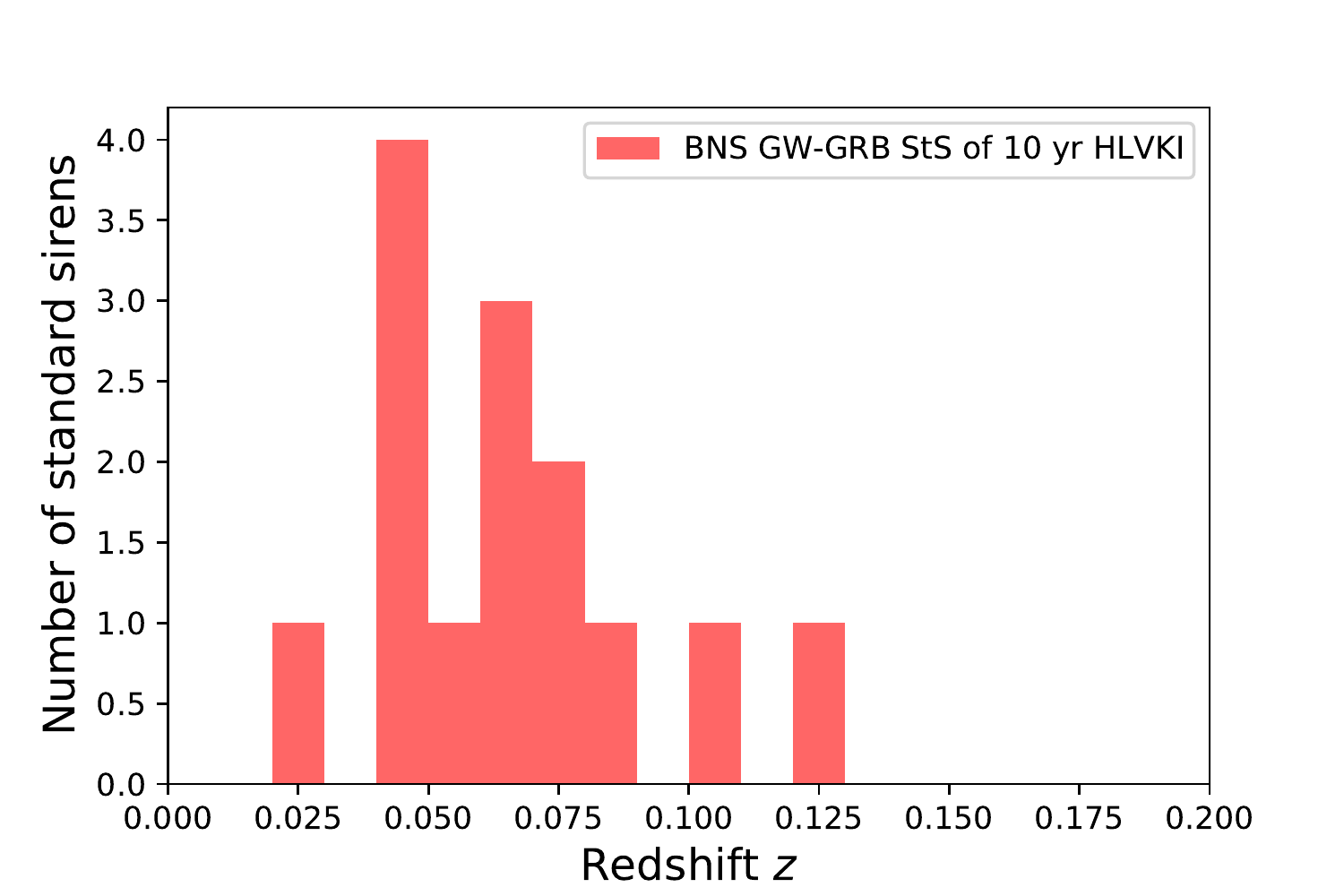}
\caption{A realization of the mock catalogue of 10 years detections of BNS GW (left) and GW-GRB standard sirens   (StS) (right) from HLVKI network (80\% duty cycle for each detector) assuming a GRB detector with the characteristics of Fermi-GBM.}
\label{fig:HLVKI_GWGRB}
\end{figure}

Figure~\ref{fig:HLVKI_GWGRB} shows one realization of the GW detections and GW-GRB coincidences for 10 years observation of HLVKI assuming a GRB detector with the characteristics of Fermi-GBM. Our simulation shows there would be of order 10-14 joint BNS GW-GRB detections in 10 years observation. The number and redshift distribution are very consistent with the estimation from~\cite{Belgacem:2019tbw}.

\begin{figure}
\centering
\includegraphics[width=0.49\textwidth]{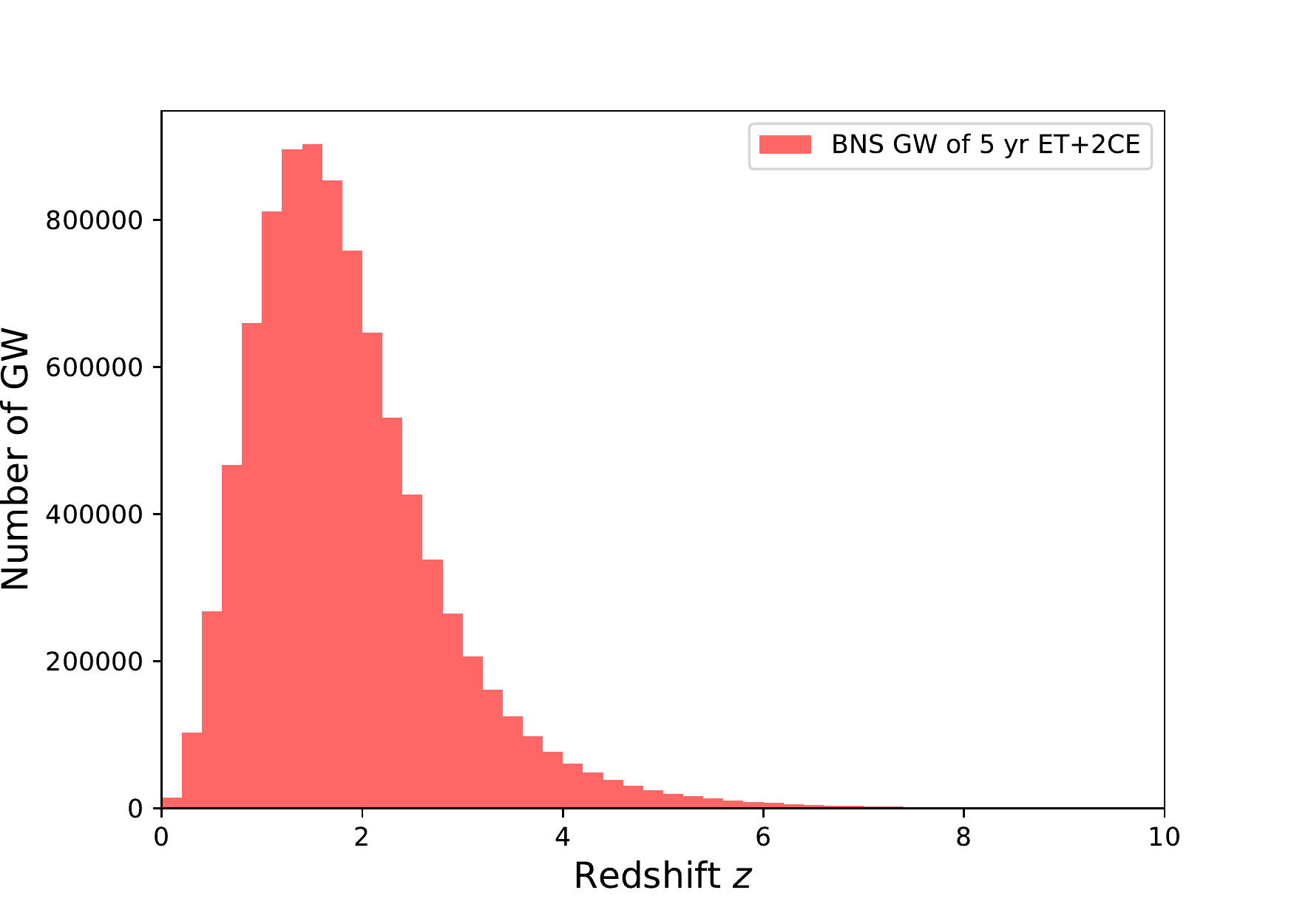}
\includegraphics[width=0.49\textwidth]{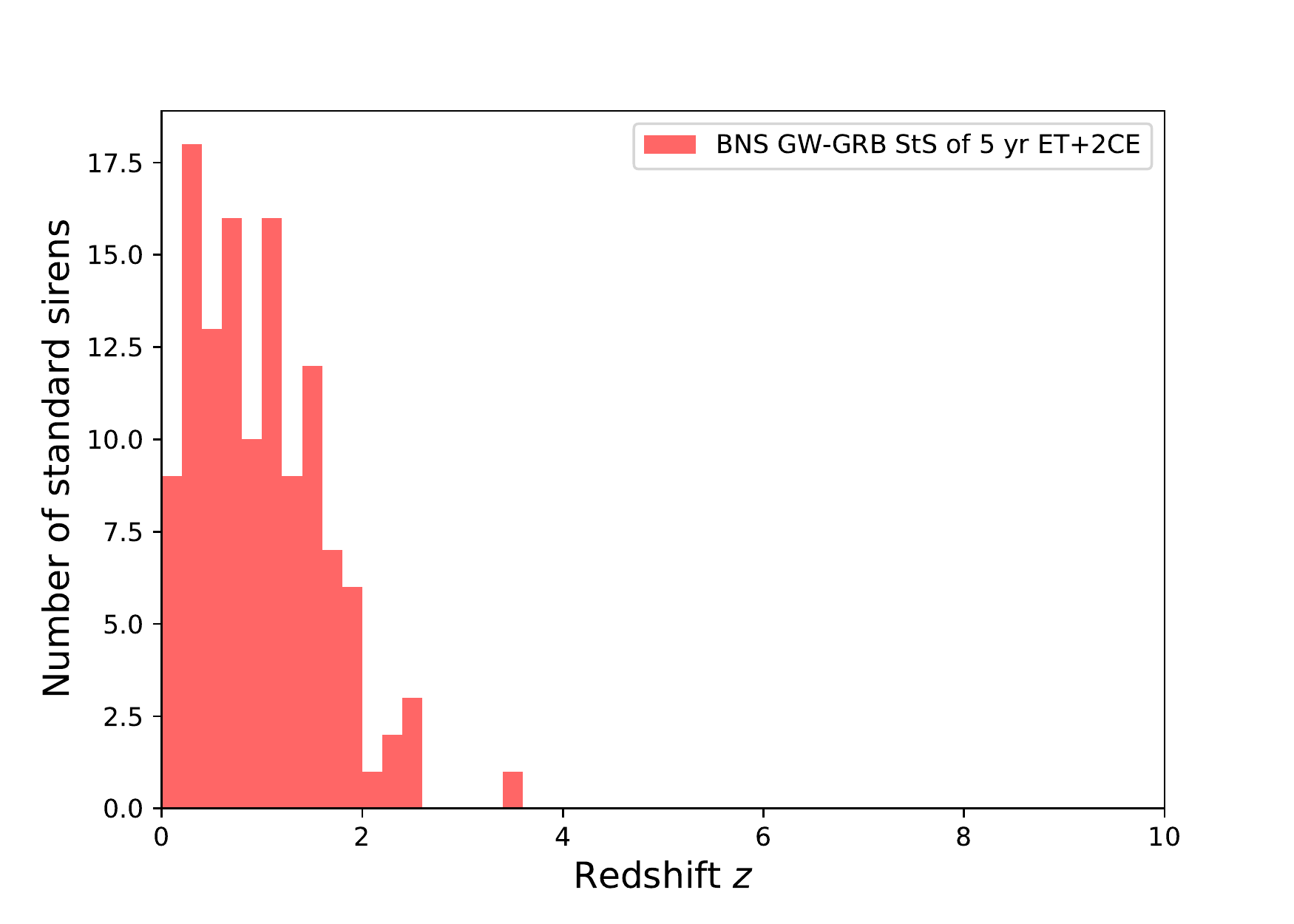}
\caption{A realization of the mock catalogue of 5 years detections of BNS GW (left) and GW-GRB standard sirens (right) from ET+2CE network (80\% duty cycle for each detector) assuming a GRB detector with the characteristics of THESEUS. We assume that only around 1/3 of the sGRBs detected by XGIS could provide redshift estimates.}
\label{fig:ET2CE_GWGRB}
\end{figure}

For the 3G network ET+2CE, the identification of the counterpart depends on the network of GRB satellites and of telescopes at the time when 3G detectors will operate. As in~\cite{Belgacem:2019tbw}, we consider THESEUS mission~\cite{Amati:2017npy,Stratta:2017bwq} to predict the coincidences between GW events and GRBs. 
For the GRB detection we assume a duty cycle of 80\% due to a reduction of 20\% as the satellite passes through the Southern Atlantic Anomaly, a flux limit of $P_{\rm T}=0.2~\rm ph~sec^{-1}~cm^{-2}$ in the 50–300 keV band and a sky coverage fraction of 0.5.
According to the THESEUS paper~\cite{Stratta:2017bwq}, only about 15-35 coincident short GRB per year will be detected by THESEUS with its X-Gamma ray Imaging Spectrometer (XGIS). We can see in figure~\ref{fig:ET2CE_GWGRB} our simulation predicts there would be of order $1.7\times10^6$ BNS detections per year by ET+2CE network, which is higher than the predictions of~\cite{Belgacem:2019tbw} for $7\times10^5$ per year and~\cite{Sathyaprakash:2019rom} for $9.9\times10^5$ per year but they are still in the same order. For the GW-GRB coincidences, our prediction is around 67 per year which is between the predictions of THESEUS paper~\cite{Stratta:2017bwq} (15-35/yr) and~\cite{Belgacem:2019tbw} (90/yr). Our redshift distribution is also very consistent with~\cite{Belgacem:2019tbw}. The XGIS will be able to localize sources to around 5 arcmin only within the central 2 sr of its field of view (FOV); outside this central region localization will be coarse at best. Following~\cite{Belgacem:2019tbw} we consider the realistic scenario, where we assume that only around 1/3 of the sGRBs detected by XGIS could provide redshift estimates. Thus the total number of GW-GRB standard sirens from 5 years observation of ET-2CE network is around 123, as shown in the right panel of figure~\ref{fig:ET2CE_GWGRB}.

Having selected the GW-GRB events one can construct the mock data of $d_L(z)$ for HLVKI and ET+2CE networks according to the fiducial cosmology and the corresponding uncertainties given above.


\subsection{MBHB standard sirens based on space-based LISA+Taiji}

To construct the mock MBHB standard sirens from the space-based LISA+Taiji network we mainly follow the strategies in~\cite{Tamanini:2016zlh,Speri:2020hwc}.  Our models for the population of MBHB are the same as those of~\cite{Klein:2015hvg}, which are in turn based on the semi-analytical galaxy formation model of~\cite{Barausse:2012fy} (see also~\cite{Sesana:2014bea,Antonini:2015cqa,Antonini:2015sza} for some improvements). Similar to~\cite{Speri:2020hwc} we choose to work with the ``popIII'' model, which assumes the ``light-seed'' scenario for the high-redshift seeds from which massive BHs grow and  produces average results with respect to the astrophysical populations considered in~\cite{Tamanini:2016zlh,Tamanini:2016uin}. The construction with the ``heavy-seed'' scenarios are straightforward. In this paper we just show the results of mock MBHB standard sirens on ``popIII'' model.

We first consider the single LISA case and then extend it to the LISA+Taiji network. We adopt the MBHB catalogues
used in~\cite{Klein:2015hvg} and sample the distribution of redshift, mass, spin parameters according to the corresponding probability densities. Since the inspiral and merger of MBHBs can last between several days to years in the frequency band of LISA and Taiji, unlike the ground-based detector case we should take into account of the motion of the detectors in space. Thus the time dependence of the detector response function plays a crucial role in localizing the position of the gravitational-wave source. To describe the coordinates of the detector we work in a heliocentric, ecliptic coordinate system. In this system the Sun is placed at the origin, the detector response functions can be written as
\begin{align}
&F^+(t)=\frac{1}{2}\left(\cos(2\psi)D^+(t)-\sin(2\psi)D^{\times}(t)\right) \,, \\
&F^{\times}(t)=\frac{1}{2}\left(\sin(2\psi)D^+(t)+\cos(2\psi)D^{\times}(t)\right) \,.
\end{align}
The full response of a space-based gravitational wave detector was complicated by the intrinsic arm-length fluctuations, pointing ahead, and the signal-cancellation accounted for in the transfer functions. As a first approximation to the response of LISA one can neglect all of these effects. So we can work to linear order in the spacecraft positions, evaluate all spacecraft locations at a common time, and set the transfer functions to unity. 
Below the transfer frequency $f_*$ the transfer functions approach unity. 
Then using the low frequency approximation $f\ll f_*$ and $f/\dot{f}\ll L$ we can have~\cite{Rubbo:2003ap}
\begin{align}
D^+(t)=&\frac{\sqrt{3}}{64}\bigg[-36\sin^2\theta\sin\big(2\alpha(t)-2\lambda\big)  \nonumber \\
            &+\big(3+\cos(2\lambda)\big)\bigg(\cos(2\phi)\Big(9\sin(2\lambda)-\sin\big(4\alpha(t)-2\lambda\big)\Big) \nonumber \\
            &+\sin(2\phi)\Big(\cos\big(4\alpha(t)-2\lambda\big)-9\cos(2\lambda)\Big)\bigg) \nonumber \\
            &-4\sqrt{3}\sin(2\theta)\Big(\sin\big(3\alpha(t)-2\lambda-\phi\big)-3\sin\big(\alpha(t)-2\lambda+\phi\big)\Big)\bigg] \,,
\end{align}
and 
\begin{align}
D^{\times}(t)=&\frac{1}{16}\Big[\sqrt{3}\cos\theta\Big(9\cos(2\phi-2\lambda)-\cos\big(4\alpha(t)-2\lambda-2\phi\big)\Big) \nonumber \\
                      &-6\sin\theta\Big(\cos\big(3\alpha(t)-2\lambda-\phi\big)+3\cos\big(\alpha(t)-2\lambda+\phi\big)\Big)\Big] \,.
\end{align}
Here $(\theta, \phi)$ is the sky location of the source, $\psi$ is the polarization angle. We sample these three parameters together with the inclination angle $\iota$ from the isotropic distribution.  $\alpha(t)=2\pi f_m t+\kappa$ is the orbital phase of the guiding center with $f_m=1/\rm yr$. The parameters $\kappa$ and $\lambda$ give the initial ecliptic longitude and orientation of the constellation. As noted in~\cite{Rubbo:2003ap} the analytical formalism for the low frequency approximation is equivalent to that derived by~\cite{Cutler:1997ta}. A single LISA can be equivalently considered as a combination of two independent detectors with the second response function is  just $F^{+,\times}(t,\theta,\phi-\pi/4,\psi)$~\cite{Cutler:1997ta,Klein:2015hvg}. The corresponding set-up for a single Taiji detector is very similar~\cite{Ruan:2019tje}.

From the sampled parameters we construct the whole catalogue of MBHBs. For each of the sampled MBHBs we can calculate the SNR from the Fourier-domain inspiral-only nonspinning waveform as in~\cite{Ruan:2019tje}. We select the GW detections with SNR>8~\cite{Klein:2015hvg,Tamanini:2016zlh,Belgacem:2019pkk}. For 5 years of observation, the number of MBHB GW events is about 370 which is very close to the results in~\cite{Tamanini:2016zlh} (see table 9 and table 10 for the configuration of ``N2A2M5L6'' which is comparable to that of LISA considered in this paper). Note~\cite{Klein:2015hvg,Tamanini:2016zlh} utilised a gravitational waveform model with generic precessing spins, called ``shifted uniform asymptotics'' (SUA) waveform, for the inspiral phase. Then they further corrected the results of their analysis to account for the effect of the merger and ringdown by using results obtained with aligned (or anti-aligned) spin inspiral-merger-ringdown (IMR) ``Phenom'' waveforms and a set of dedicated precessing-spin IMR hybrid waveforms (see the references for details). This correction becomes increasingly significant for heavier MBHBs (we can see for the popIII model there is only a subtle improvement when including the merger and ringdown). However, since we just focus on the light-seed MBHBs model in this paper we can simplify our analysis based on the inspiral-only nonspinning waveform. We find our result is very consistent with either conservative scenario (inspiral only) or optimistic scenario (with merger and ringdown) in~\cite{Tamanini:2016zlh}. 

Among all the LISA detections with SNR> 8 we should select the events with a sky localization $\Delta \Omega<10~ \rm deg^2$ as the potential GW events with EM counterparts (corresponding to the field of view of such as the LSST survey). Using the Fisher matrix $\Gamma_{ij}=\left(\frac{\partial h}{\partial \theta_i}\Big{|}\frac{\partial h}{\partial \theta_j}\right)$ one can estimate the measured errors of the parameter $\theta_i$. While in this paper we would like to use the results of~\cite{Tamanini:2016zlh} as an anchor to estimate the sky localization $\Delta \Omega$. We know the instrument error of luminosity distance is proportional to SNR$^{-1}$. The analysis in~\cite{Klein:2015hvg} shows $\Delta d_L/d_L\propto \mathcal{R}^{-1}$ and $\Delta \Omega \propto \mathcal{R}^{-2}$. Here $\mathcal{R}$ is SNR gain when accounted for the merger and ringdown. Thus we can directly use the number of cases with $\Delta \Omega<10~\rm deg^2$ in table 10 of~\cite{Tamanini:2016zlh} (here we adopt the optimistic scenario for ``N2A2M5L6'' which gives the number 35) as an anchor to estimate the distribution of the sky locations for all of the MBHB GW events estimated by a single LISA detector.

Following~\cite{Tamanini:2016zlh} we characterize the EM emission at merger by assuming the production of an optical accretion-powered luminosity flare and of radio flares and jets, based on results from general-relativistic simulations of merging MBHBs in an external magnetic field~\cite{Palenzuela:2010nf}. We use the simulations of MBHB catalogs to compute the magnitude of the EM emission of each MBHB GW event in both the optical and radio bands and thus determine the number of counterparts detected by future EM facilities, specifically LSST~\footnote{\url{www.lsst.org.}}, SKA~\footnote{\url{www.skatelescope.org.}} and ELT~\footnote{\url{www.eso.org/sci/facilities/eelt/.}}. As shown by~\cite{Tamanini:2016zlh}, in fact the counterparts detectable by LSST are always detectable by SKA+ELT. The number of LSST counterpart detections is around 1.  Thus in this paper we just calculate the cases for SKA and ELT. From our calculation, within 5 years observation of MBHB GW events by LISA, the number of the radio counterparts observed by SKA is 34; the number of the optical observations with ELT of SKA counterparts hosts is 28. In more detail, the number of counterparts whose redshift can be measured spectroscopically (photometrically) is 12 (16). All of our estimations of EM counterparts are very consistent with the numbers given by~\cite{Tamanini:2016zlh}.

Now we extend our analysis to the LISA+Taiji network. The improvement of the joint LISA+Taiji for standard sirens should be accounted for two aspects. First the joint of Taiji would improve the SNR of the MBHB GW signal. Thus more cases would be detected. Our simulation shows there will be about 414 MBHB GW cases compared to 370 cases for the single LISA. The error of the inferred luminosity distance would be improved accordingly. 
Actually the instrumental error of $d_L$ is very subdominant compared to other errors such as from the lensing~\cite{Tamanini:2016zlh,Tamanini:2016uin,Speri:2020hwc}. Thus the reduction of the instrumental error of $d_L$ would only subtly improve the performance of measuring the luminosity distance . In this paper we follow~\cite{Speri:2020hwc} to estimate the LISA instrumental error $\Delta d_L/d_L=0.05(d_L/(d_L(4)~\rm Gpc))$, which is anchored from the recent results by the full Bayesian approach in~\cite{Marsat:2020rtl}. The second but more important aspect of the joint LISA+Taiji for standard sirens is the significant improvement of the localization of the GW sources. For example,~\cite{Ruan:2019tje,Ruan:2020smc} have shown that for an equal-mass black hole binary located at redshift of 1 with a total intrinsic mass of $10^5 M_\odot$, the LISA-Taiji network may achieves about 4 orders of magnitude improvement on the event localization region compared to an individual detector (that is 3 orders of magnitude for $\Delta \Omega$ and 1 order of magnitude for $d_L$ ). Recently,~\cite{Wang:2020vkg} showed the angular resolution of LISA+Taiji network can be improved by more than 10 times comparing the single detector. In this paper, based on the previous research, we assign an average improvement of the localization $\Delta \Omega$ to be 10 times which is a conservative estimation. From our calculation, for 5 years of observation of LISA+Taiji network, the number of SKA+ELT counterparts is around  58 (if the average improvement of localization is 100 times the number is 83). Considering the realistic overlap of the operation time between LISA and Taiji in the 2030s, we think this number is a reasonable estimation. In a full analysis of LISA+Taiji network, one should do the Bayesian parameter estimation for every MBHB mergers in the simulated catalogue and take into account of the whole processing inspiral, merger and ringdown phases. We would like to leave this for future research.

\begin{figure}
\centering
\includegraphics[width=0.49\textwidth]{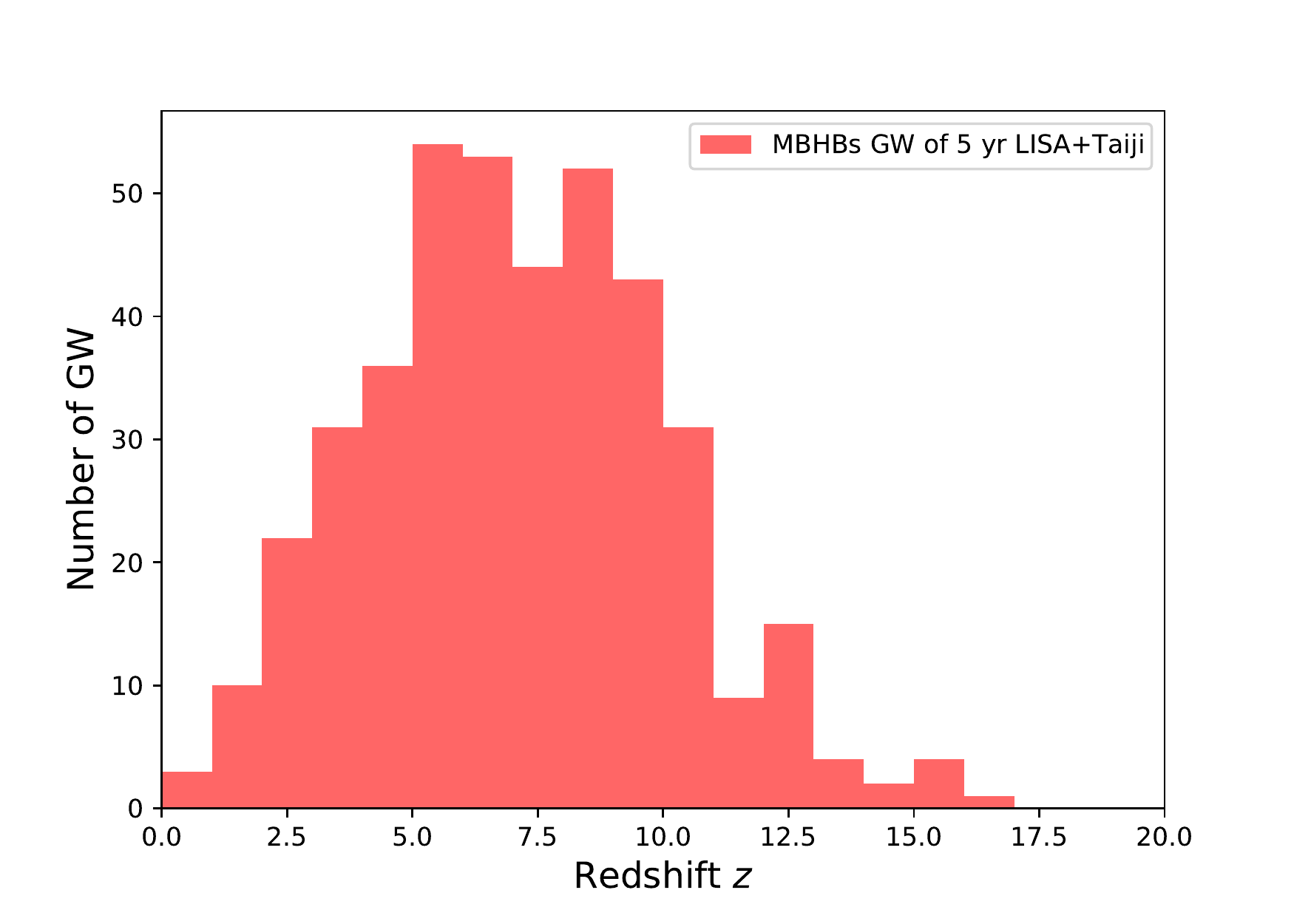}
\includegraphics[width=0.49\textwidth]{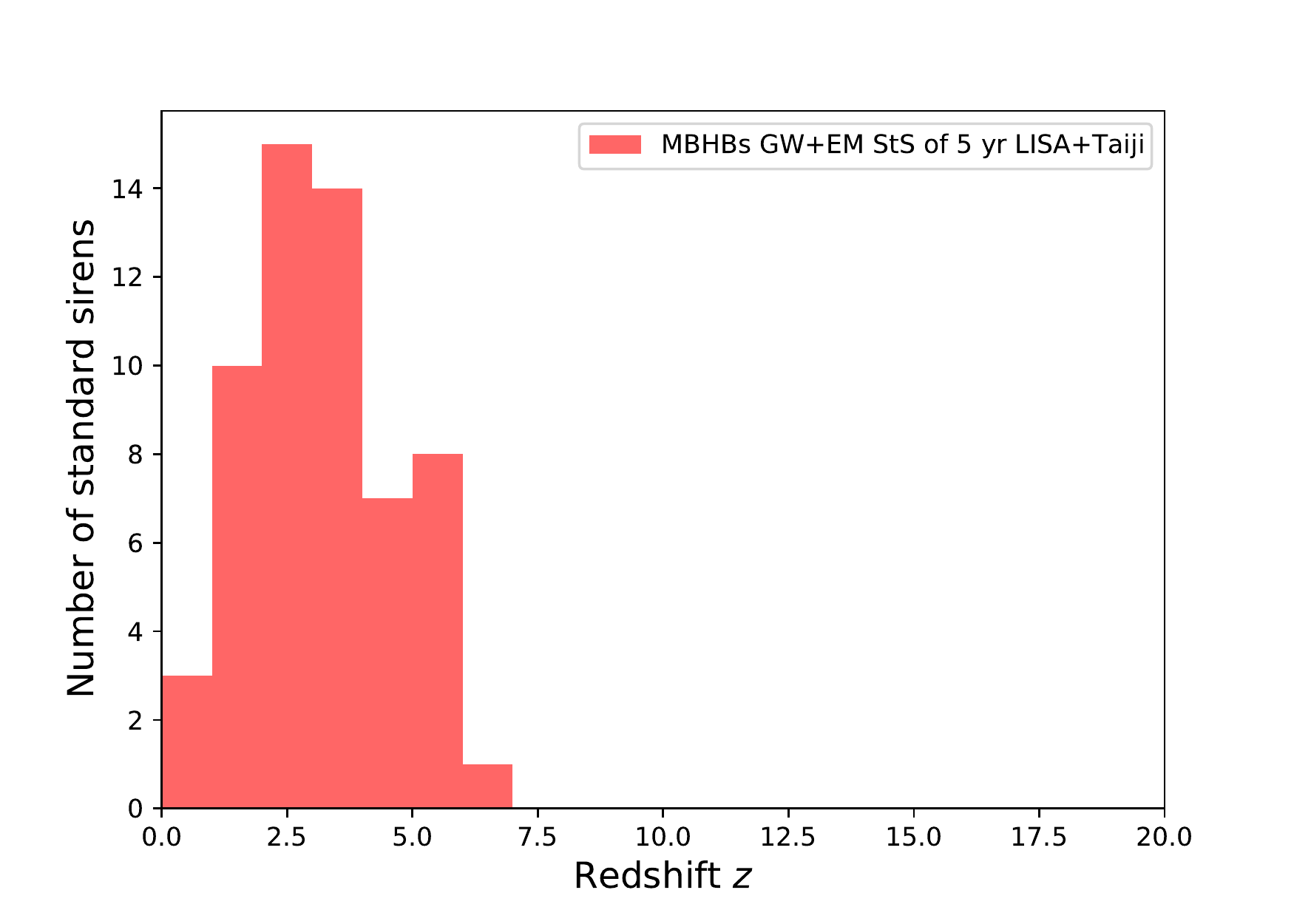}
\caption{A realization of the mock catalogue of the 5 years detections of MBHBs GW (left) and MBHBs GW+EM standard sirens (right) from LISA+Taiji network. We calculate specifically the number of counterparts whose redshift can be measured spectroscopically and photometrically from ELT. }
\label{fig:LISA_TAIJI_GW}
\end{figure}

Figure~\ref{fig:LISA_TAIJI_GW} shows our constructed catalogue of MBHB standard sirens for LISA+Taiji network. 
We can see the network can detect the GW emitted from the merger of MBHBs all the way back to their earliest formation at redshift around $z\sim16$. Since the limitation by the observation of EM counterpart, the useful standard sirens redshifts can only reach around 6--7. Now it is straightforward to sample the $d_L-z$ data from the total uncertainty of $d_L$ in the fiducial cosmology. Note we calculate specifically the number of counterparts whose redshift can be measured spectroscopically and photometrically from ELT. We assign an additional error of redshift $\Delta z=0.03(1+z)$ for the photometric measurements of ELT~\cite{Kruhler:2010jw,Dahlen:2013fea}. We propagate this redshift uncertainty to the distance uncertainty in the fiducial cosmology.

\subsection{The combined Hubble diagram of standard sirens}

\begin{figure}
\centering
\includegraphics[width=0.9\textwidth]{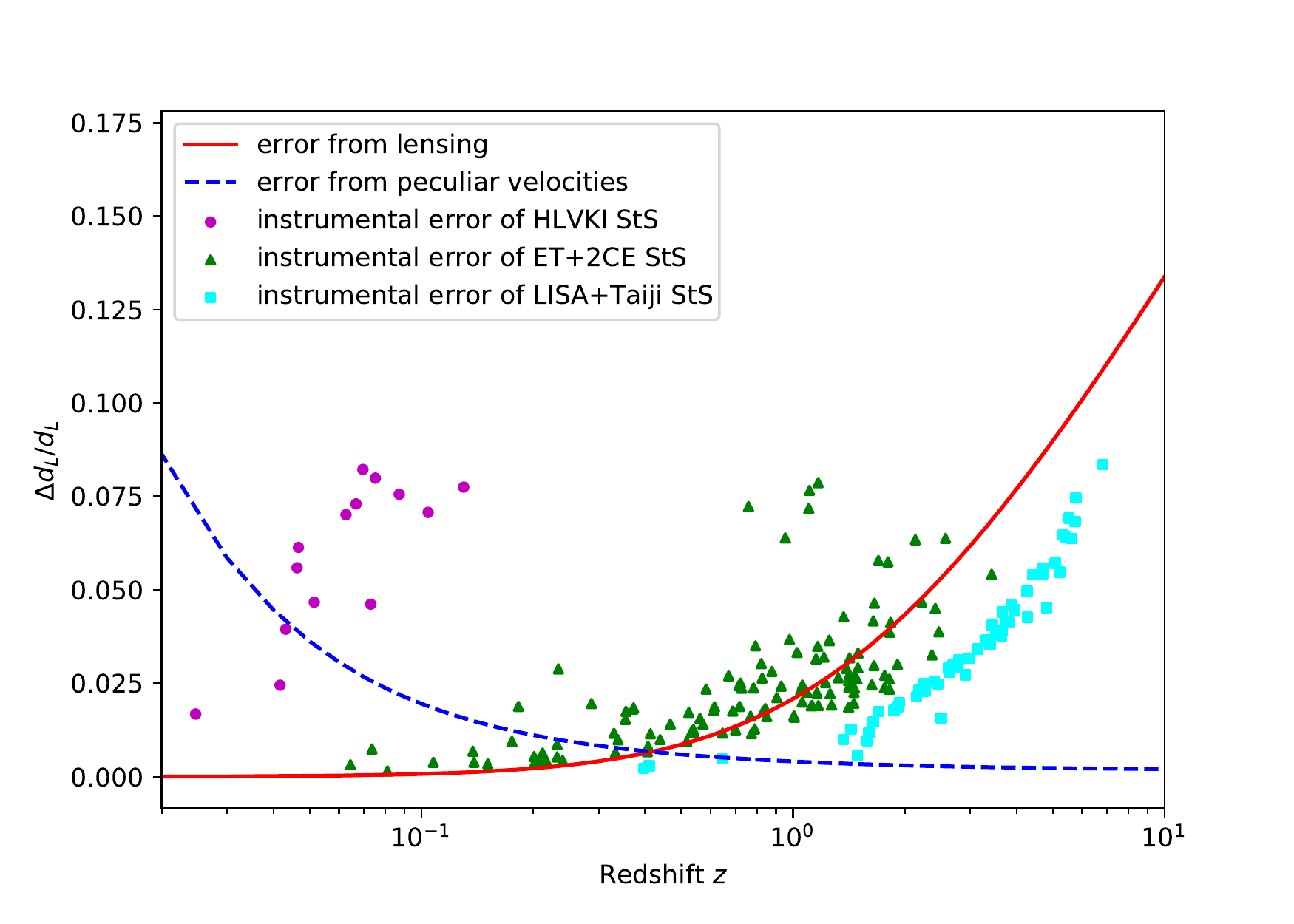}
\caption{The relative error of luminosity distance from lensing (with delensing), peculiar velocities, and the instrumental error of the sampled standard sirens by different networks.   }
\label{fig:error_dL}
\end{figure}

\begin{figure}
\centering
\includegraphics[width=0.9\textwidth]{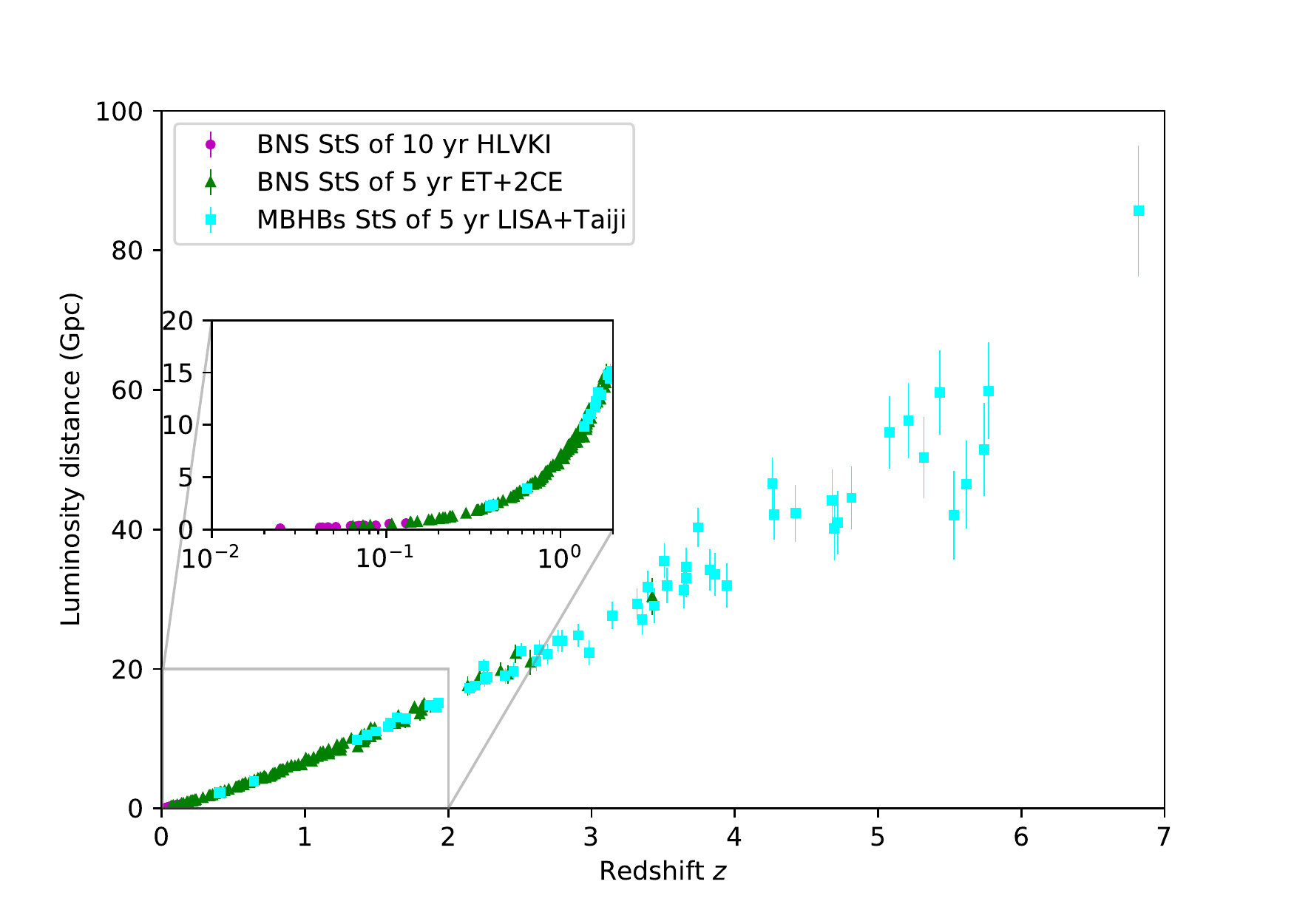}
\caption{The Hubble diagram of one realization of mock standard sirens from future GW detect networks. }
\label{fig:Hubble_diagram}
\end{figure}

We summarize our estimation of the numbers of GW and GW+EM events for future GW detector networks in table~\ref{tab:numbers}.  
From our simulations above we can see the redshift distributions of standard sirens for different GW detector networks, i.e., $z$ mainly falls in [0,0.15] for HLVKI, [0,3] for ET+2CE, and [1,6] for LISA+Taiji.  To construct the expansion history of Universe, intuitively we can combine the standard sirens catalogues of these networks together. Figure~\ref{fig:error_dL} shows the relative errors of luminosity distanced from different contributions. For HLVKI, the error of $d_L$ is dominated by the instrumental one and the peculiar velocities also play important roles. While for ET+2CE case, both lensing and instrumental error count and the latter is relatively larger than former. Finally for the LISA+Taiji case, we can see clearly the error is dominated by the lensing. 
By combining these uncertainties together for every standard sirens, we can draw the measured Hubble diagram by these future GW detector networks. The Hubble diagram of standard sirens from the future GW detector networks that we estimate would be released in the 2030s is shown in figure~\ref{fig:Hubble_diagram}. In this paper, we would like to use the combined Hubble diagram as an overall estimation of the potential of GW standard sirens on studying cosmology and modified gravity theory in the 2030s.

\begin{table}
\centering
\begin{tabular}{l|cc} 
\hline\hline
Network           & GW events                   & Joint GW+EM events  \\ 
\hline
HLVKI (10 yr)     & 865 (BNS)     & 14 (Fermi-GBM)      \\
ET+2CE (5 yr)     & 8929810 (BNS) & 123 (THESEUS)       \\
LISA+Taiji (5 yr) & 414 (MBHB)     & 58 (SKA+ELT)        \\
\hline
\end{tabular}
\caption{The estimation of the numbers of GW and GW+EM events for future GW detector networks. The parenthesis in each column represents (from left to right): operation time for each network, source type of the GW events, the corresponding EM counterpart detectors. Note the numbers of the joint GW+EM events only include that have the redshift measurements. }
\label{tab:numbers}
\end{table}

\section{Standard sirens on cosmology and modified gravity theory \label{sec:StSCosMG}}

Having the combined Hubble diagram of standard sirens from these networks  it is very straightforward to use them to constrain the cosmological parameters. Note the $d_L-z$ relation of standard sirens does not rely on the calibration like SNe, they can be directly used to constrain the parameters which is embedded in the luminosity distance, such as the Hubble constant. We start from the concordance model, namely the baseline $\Lambda$CDM, in which dark energy is just a cosmological constant with equation of state $w=-1$. We would like to forecast how precisely the future standard sirens would be helpful to constrain the Hubble constant and matter density parameter. Then we extend the base model to study the dynamics of dark energy. Like the usual approaches in the literature we consider two ways of the parameterization of the equation of state of dark energy, i.e., the $w$CDM and $w_0w_a$CDM. The former assumes a constant equation of state $w$ while the latter parameterizes $w$ in the Chevallier-Polarski-Linder (CPL) form as $w(z)=w_0+w_az/(1+z)$~\cite{Chevallier:2000qy}. Finally we study the modified propagation of GW with a phenomenological parameterization of the modified GW luminosity distance to constrain the MG theories.


\subsection{Explicit model-fitting MCMC approach \label{sec:MCMC}}


\subsubsection{Base $\Lambda$CDM and its extensions}

To select the mock standard sirens of GW detector networks, we construct 30 realizations of cataloguess of the Hubble diagram for HLVKI, ET+2CE and LISA+Taiji using different random seeds. For each of the realizations we use Gaussian process to reconstruct $d_L(z)$ and its derivatives from the mock data (the details of GP and its applications to standard sirens will be shown in section~\ref{sec:ML}).  From these reconstructed functions we select a representative catalogue, from which the reconstructed mean functions of $d_L$ and its derivatives are consistent with the fiducial cosmology as we set for the simulation. The reason for this selection strategy is as follows.  The different realizations of the mock data sets only differ in the mean value of the measurement, while the statistic information (error, variance) maintain the same. From the simulation, we mainly focus on how precisely we can constrain the parameters., i.e, the covariance of the posteriors.  The absolute best-fit value is less important due to the scattering in the process of simulation.  We shall choose the one which gives consistent GP reconstructions of luminosity distance (relative to the fiducial model) in section~\ref{sec:ML} for a better illustration. To make the catalogues uniformly adopted throughout this paper, we select the catalogue based on the GP reconstructions. However, we have checked that different realizations of catalogues would not change our results. We use ``StS'' to denote the combined standard sirens catalogues of HLVKI, ET+2CE, and LISA+Taiji that we select to study cosmology and modified gravity theory below.

\begin{figure}
\centering
\includegraphics[width=0.8\textwidth]{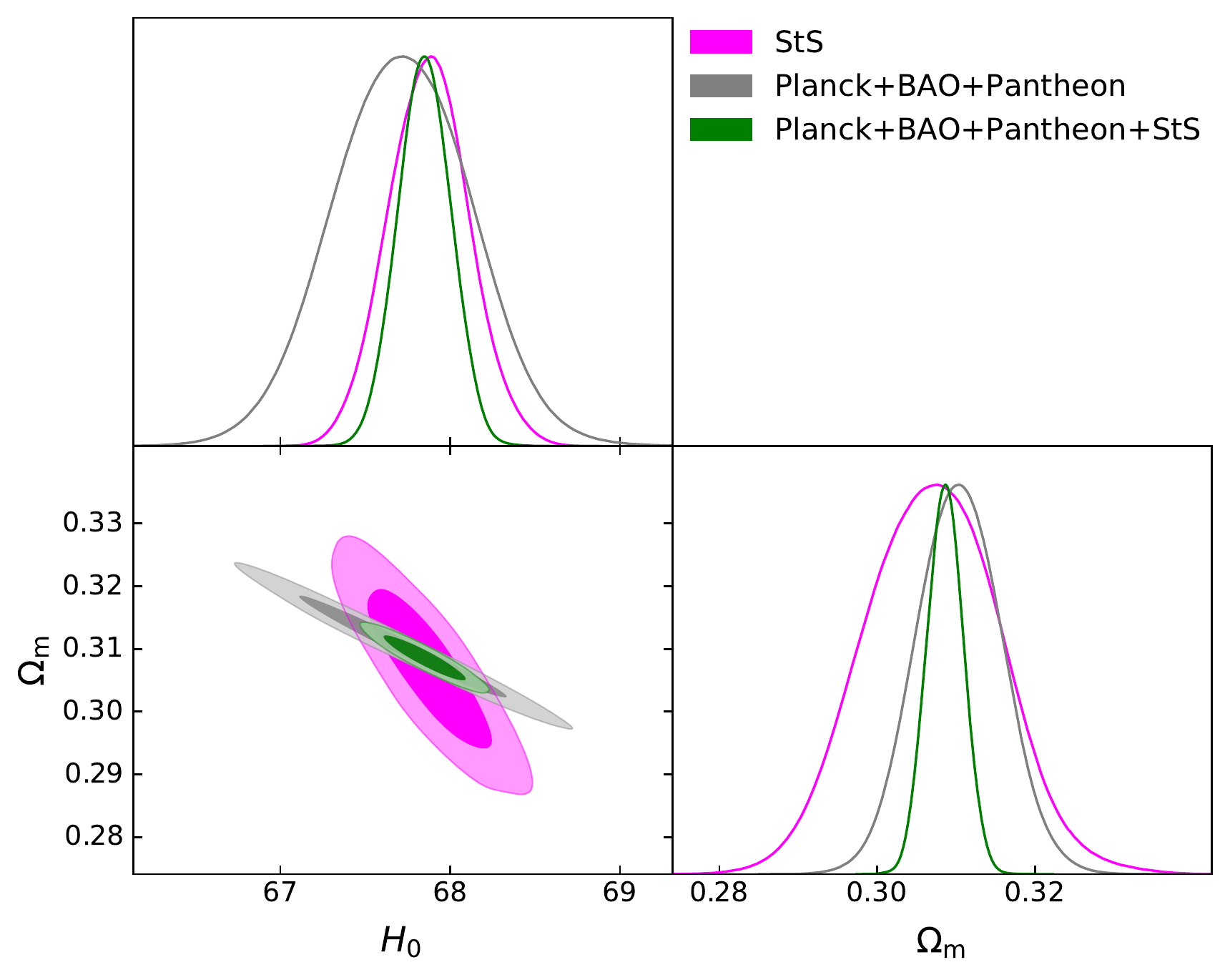}
\caption{Constraints on parameters of the base $\Lambda$CDM model from the combined mock standard sirens of future GW detector networks. We also plot the current {\it Planck}+BAO+Pantheon results as a comparison. Contours contain 68 \% and 95 \% of the probability.}
\label{fig:LCDM_MCMC}
\end{figure}

\begin{figure}
\centering
\includegraphics[width=0.8\textwidth]{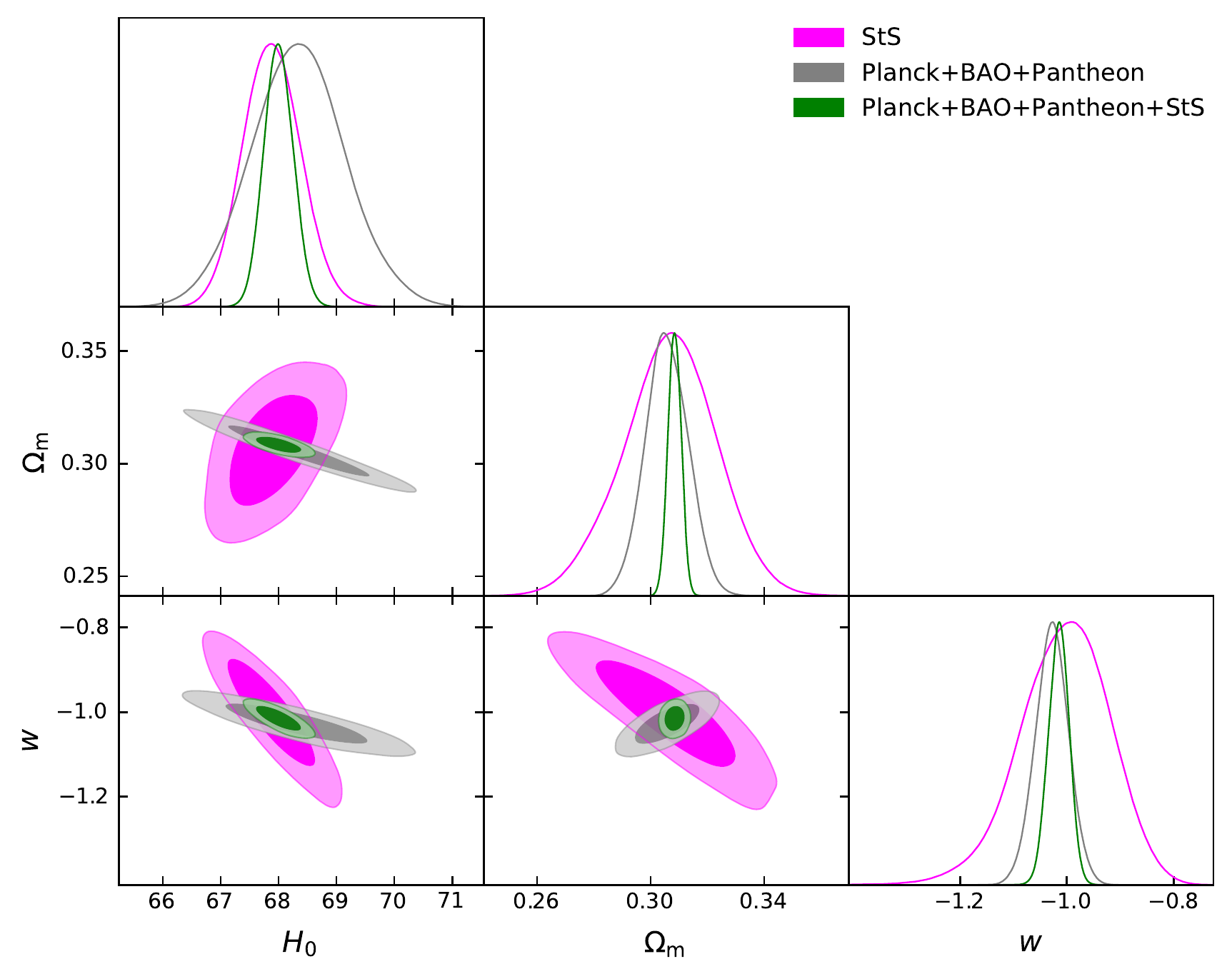}
\caption{Constraints on parameters of the $w$CDM model from the combined mock standard sirens of future GW detector networks.}
\label{fig:wCDM_MCMC}
\end{figure}

\begin{figure}
\centering
\includegraphics[width=0.8\textwidth]{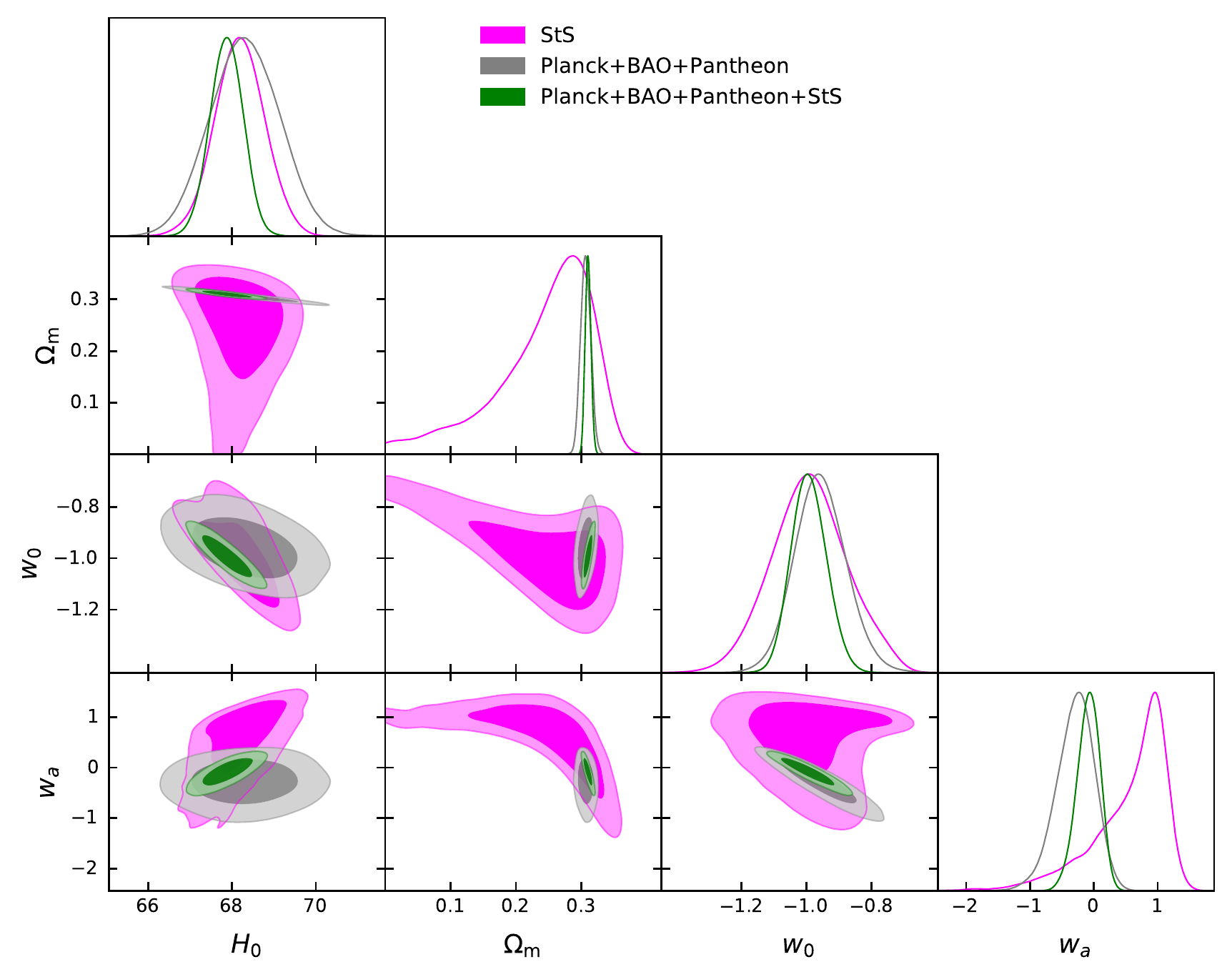}
\caption{Constraints on parameters of the CPL model from the combined mock standard sirens of future GW detector networks.}
\label{fig:CPL_MCMC}
\end{figure}

We first constrain the base $\Lambda$CDM model from the combined Hubble diagram of StS. We use the MCMC package {\sc Cobaya}~\cite{Torrado:2020dgo,2019ascl.soft10019T} to explore the parameter space and obtain the posteriors. We also use {\sc CosmoMC}~\footnote{\url{https://cosmologist.info/cosmomc/}}~\cite{Lewis:2002ah} as a cross check to make sure the results of these two packages are consistent with each other. The marginalized statistics of the parameters and the plots are produced by the Python package {\sc GetDist}~\cite{Lewis:2019xzd}. Figure~\ref{fig:LCDM_MCMC} shows the constraints of Hubble constant and matter density parameter in $\Lambda$CDM. For comparison, we also include the traditional EM experiments such as CMB, BAO and SNe Ia. In this paper, we use the CMB data from latest {\it Planck}~\cite{Aghanim:2018eyx}, that is, \textit{Planck} TT,TE,EE+lowE+lensing (briefly, we just write it as {\it Planck} hereafter).  For BAO we adopt the isotropic constraints provided by 6dFGS at $z_{\rm eff}=0.106$~\cite{Beutler:2011hx}, SDSS-MGS DR7 at $z_{\rm eff}=0.15$~\cite{Ross:2014qpa}, and ``consensus'' BAOs in three redshift slices with effective redshifts $z_{\rm eff}$ = 0.38, 0.51, and 0.61~\cite{Ross:2016gvb,Vargas-Magana:2016imr,Beutler:2016ixs}. We use the Pantheon data~\cite{Scolnic:2017caz} as the latest compilation of SNe Ia. We show the results from three data combinations as comparison. The first is the standard sirens alone. The second is \textit{Planck}+BAO+Pantheon data combination which we obtain from Planck Legacy Archive~\footnote{\url{http://pla.esac.esa.int/}}. Finally we combine all of them together. Similarly, the constraints of $w$CDM and CPL model are shown in figures~\ref{fig:wCDM_MCMC} and~\ref{fig:CPL_MCMC}.

\subsubsection{Phenomenological parametrization of modified GW propagation \label{sec:MGGW}}

Recent studies have shown that GR can be tested by the propagation of GWs across cosmological distances~\cite{Belgacem:2017ihm,Belgacem:2018lbp,Nishizawa:2017nef,Arai:2017hxj,Nishizawa:2019rra,Belgacem:2019pkk,Belgacem:2019tbw,Belgacem:2019zzu,Mukherjee:2019wcg,DAgostino:2019hvh,Bonilla:2019mbm,Mukherjee:2020mha,Kalomenopoulos:2020klp,Mastrogiovanni:2020mvm,Mastrogiovanni:2020gua,Finke:2021aom}. In a generic modified gravity model the linearised evolution equation for GWs traveling on an FRW background in four dimensional space-time is~\cite{Belgacem:2019pkk}
\begin{equation}
\tilde{h}_A^{\prime\prime}+2[1-\delta(\eta)]\mathcal{H}\tilde{h}_A^{\prime}+[c_T^2(\eta)k^2+m_T^2(\eta)]\tilde{h}_A=\Pi_A \,,
\label{eq:GWpropa}
\end{equation}
where $\tilde{h}_A$ are the Fourier modes of the GW amplitude. $\mathcal{H}=a^{\prime}/a$ is the Hubble parameter in conformal time, the primes indicate derivatives with respect to conformal time $\eta$, $A=+,\times$ labels the two polarizations, and $\Pi_A$ is the source term, related to the anisotropic stress tensor. The function $\delta (\eta)$ modifies the frication term in the propagation equation. $c_T$ corresponds to the speed of gravitational waves.  In theories of modified gravity the tensor mode can be massive, with $m_T$ its mass. In GR we have $\delta=0$, $c_T=c$,  and $m_T=0$. The observation of GW170817/GRB170817A put a very tight constraint of the speed of gravitational wave, $(c_T-c)/c<\mathcal{O}(10^{-15})$~\cite{Monitor:2017mdv}. In this paper, Following~\cite{Belgacem:2018lbp,Belgacem:2019tbw} we only retain the deviations from GR induced by the friction term,
\begin{equation}
h_A^{\prime\prime}+2[1-\delta(\eta)]\mathcal{H}h_A^{\prime}+k^2h_A=0 \,.
\label{eq:GWpropa_friction}
\end{equation}
Then one can show the inferred ``GW luminosity distance'' in modified gravity theories is different from the traditional ``electromagnetic luminosity distance''~\cite{Belgacem:2017ihm,Belgacem:2018lbp},
\begin{equation}
d_L^{\rm gw}(z)=d_L^{\rm em}(z)\exp\left\{-\int_0^z\frac{dz^{\prime}}{1+z^{\prime}}\delta(z^{\prime})\right\} \,.
\end{equation}
To constrain the modified gravity theory (or to test GR), we need to constrain the $\delta$ function in~\eqref{eq:GWpropa_friction}. In this section, we follow~\cite{Belgacem:2018lbp,Belgacem:2019tbw} to adopt the 2-parameter phenomenological parameterization
\begin{equation}
\Xi(z)\equiv\frac{d_L^{\rm gw}(z)}{d_L^{\rm em}(z)}=\Xi_0+\frac{1-\Xi_0}{(1+z)^n} \,.
\label{eq:Xi}
\end{equation}
GW measurements can therefore access the quantity $\delta(z)$, or equivalently $\Xi(z)$, a smoking gun of modified gravity. Obviously, $\Xi_0=1$ in GR.
This parametrization was originally proposed in~\cite{Belgacem:2018lbp}, inspired by the fact that it fits extremely well the prediction for $\Xi(z)$ obtained from a nonlocal modification of gravity~\cite{Belgacem:2017cqo}, but it was then realized that its features are very general, so that it is expected to fit the predictions from a large class of models (for the explicit predictions of modified gravity models such as the scalar-tensor theories of the Horndeski class in terms of this parametrization please refer to~\cite{Belgacem:2019pkk} and references therein). In several explicit MG models,  we can also obtain a corresponding parametrization for the time variation of the effective Planck mass or of the effective Newton constant~\cite{Belgacem:2019pkk},
\begin{align}
&M_{\rm eff}(z)=M_{\rm pl}\Xi^{-1}(z) \,, \label{eq:Meff}\\
&G_{\rm eff}(z)=G\Xi^2(z) \,. \label{eq:Geff}
\end{align}
Thus the constraints of $\Xi(z)$ can be interpreted as the null test of the variation of effective Planck mass or Newton constant in some specific modified gravity theories. However, in this paper we only focus on free parameter $\Xi(z)$ as a smoking gun of the modified gravity effects.

\begin{figure}
\centering
\includegraphics[width=0.8\textwidth]{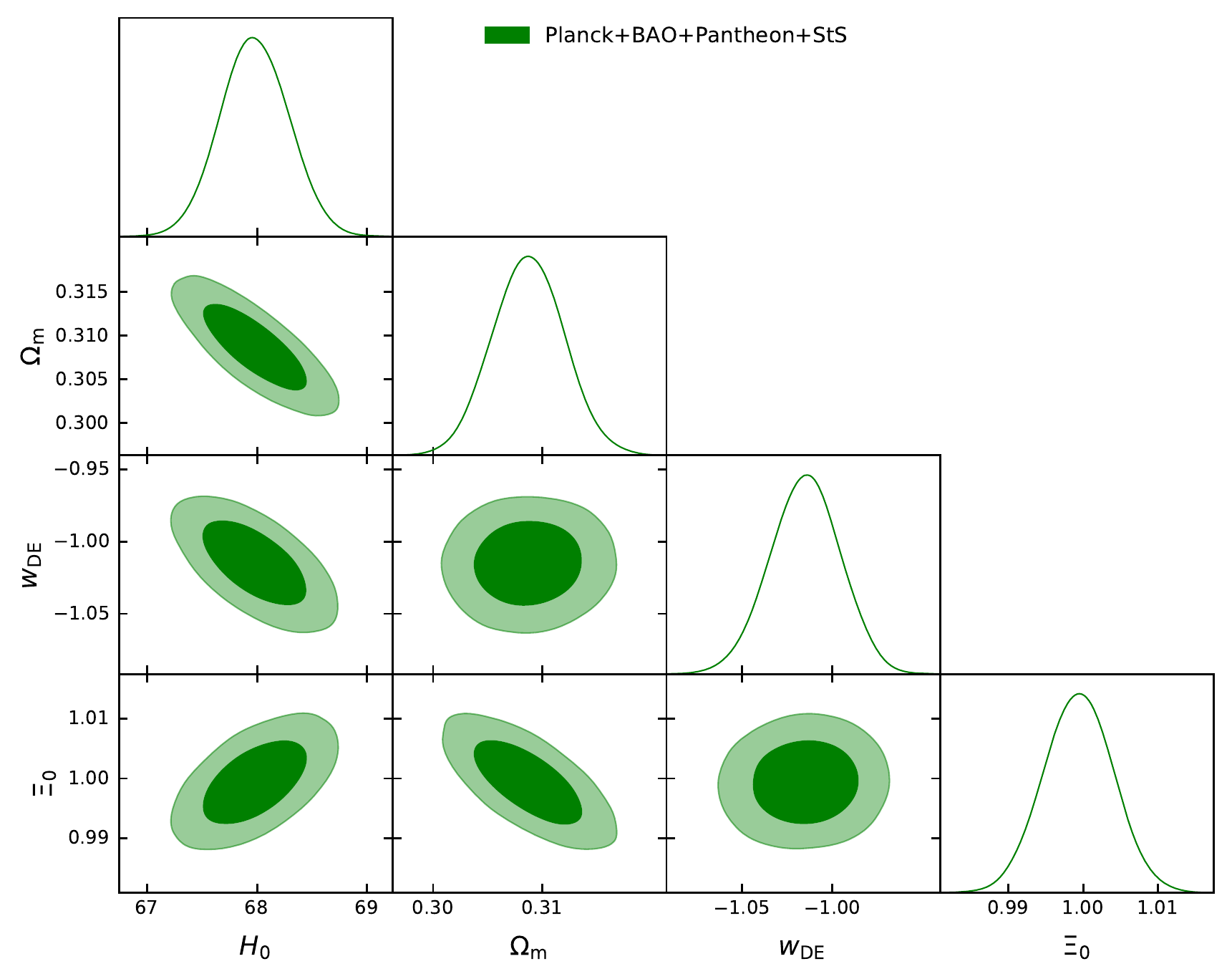}
\caption{Constraints on parameters of a phenomenological parameterization of modified GW propagation from the combined mock standard sirens of future GW detector networks together with {\it Planck}, BAO and Pantheon.}
\label{fig:MG_MCMC}
\end{figure}

Now following~\cite{Belgacem:2018lbp,Belgacem:2019tbw,Belgacem:2019pkk} we use the {\it Planck} data with BAO and Pantheon as the anchor of $d_L^{\rm em}$ and then combine the standard sirens of future GW detector networks to constrain the MG parameter $\Xi_0$ together with the dark energy equation of state $w$. We modify the package {\sc Cobaya} to incorporate the modified GW propagation effect into the codes. We set $n=2.5$ since it plays in general a lesser role as in~\cite{Belgacem:2018lbp}. The result is shown in figure~\ref{fig:MG_MCMC}.

\subsection{Machine-learning nonparametric reconstruction approaches \label{sec:ML}}

In the last decade, several model-independent or nonparametric approaches in cosmological data analysis have been proposed and studied. One of the most prominent techniques is the Gaussian process (see e.g.~\cite{Seikel:2012uu,Shafieloo:2012ht}). GP is a Bayesian regression method which belongs to the supervised machine learning. It is a purely data-driven reconstruction technique from which we can reconstruct the parameter as a function without assuming the specific form of parameterization and even any cosmological model, thus one can avoid the model bias. GP has been widely used to study cosmology with the traditional EM experiments~\cite{Cai:2015zoa,Cai:2015pia,Cai:2016vmn,Gomez-Valent:2018hwc,Cai:2019bdh,Yang:2020jze} and GW data sets~\cite{Cai:2016sby,Cai:2017yww,Cai:2017plb,Belgacem:2019zzu}. 
In the meanwhile, the Neural Networks in machine learning (deep learning) have been rapidly developed in recent years. For example, the convolutional neural networks (CNNs) are applied to the inference of cosmological parameters and models
~\cite{Schmelzle:2017vwd,PerreaultLevasseur:2017ltk,Peel:2018aei,Fluri:2019qtp}
and also the detection and data analysis of GW
~\cite{Zevin:2016qwy,George:2017pmj,Gabbard:2017lja}.
A new nonparametric approach for reconstructing a function from observational data using an Artificial Neural Network (ANN) has also been proposed and tested with the Hubble Parameter and SNe Ia~\cite{Wang:2019vxv}. In this paper, though we mainly focus on the GP regression technique, we also show the reconstruction results with ANN just as a comparison with GP.

Having the standard sirens catalogues from future GW detector networks, the nonparametric (model-independent) constraints on cosmology and modified gravity theory should be forecasted. 
We first would like to use GW standard sirens alone to reconstruct the Hubble parameter $H(z)$ and the equation of state of dark energy $w(z)$ under GR. These can be regarded as the tests of the base $\Lambda$CDM model along the redshift (back to an earlier time of the Universe) by the nonparametric approaches.  In a flat Universe with the Friedmann–Lema\^itre–Robertson–Walker (FRLW) metric, the Hubble parameter can be written as,
\begin{equation}
H=\frac{c(1+z)^2}{d_L^{\prime}(1+z)-d_L} \,,
\label{eq:Hz}
\end{equation} 
where the prime denotes the derivative with respect to redshift.
Similarly from the Friedmann equation one can write the equation of state of dark energy in terms of the luminosity distances,
\begin{equation}
w=-\frac{c_0^2(1+z)\Big(d_L+(1+z)\big(-d_L^{\prime}+2d_L^{\prime\prime}(1+z)\big)\Big)}{3\big(d_L-d_L^{\prime}(1+z)\big)\Big(-c_0^2(1+z)+\Omega_m\big(d_L-d_L^{\prime}(1+z)\big)^2\Big)} \,,
\label{eq:wz}
\end{equation}  
here $c_0=c/H_0$. From GP one can reconstruct the $d_L(z)$ function and its derivatives from the data sets. We use the {\sc GaPP} codes~\cite{Seikel:2012uu} with several improvements for GP reconstructions. We adopt the same mock standard sirens as in section~\ref{sec:MCMC}, which is selected from 30 realizations to represent a stable and consistent reconstruction (for the mean value) of the fiducial model (we have argued that this selection will not bias our results.).  Combining the covariance between the reconstructed $d_L(z)$, $d_L^{\prime}(z)$, and $d_L^{\prime\prime}(z)$,  from~\eqref{eq:Hz} we can reconstruct the Hubble parameter without assuming the specific cosmological model and from~\eqref{eq:wz} the equation of state $w(z)$ can also be reconstructed as a function of redshift and we do not need to assume the parametric form such as CPL.  Note the reconstruction of $w$ relies on the information of $H_0$ and $\Omega_m$. Since we focus on the feature of the evolving equation of state, we assume $H_0$ and $\Omega_m$ have been constrained and fix them to be the fiducial values.

\begin{figure}
\centering
\includegraphics[width=0.8\textwidth]{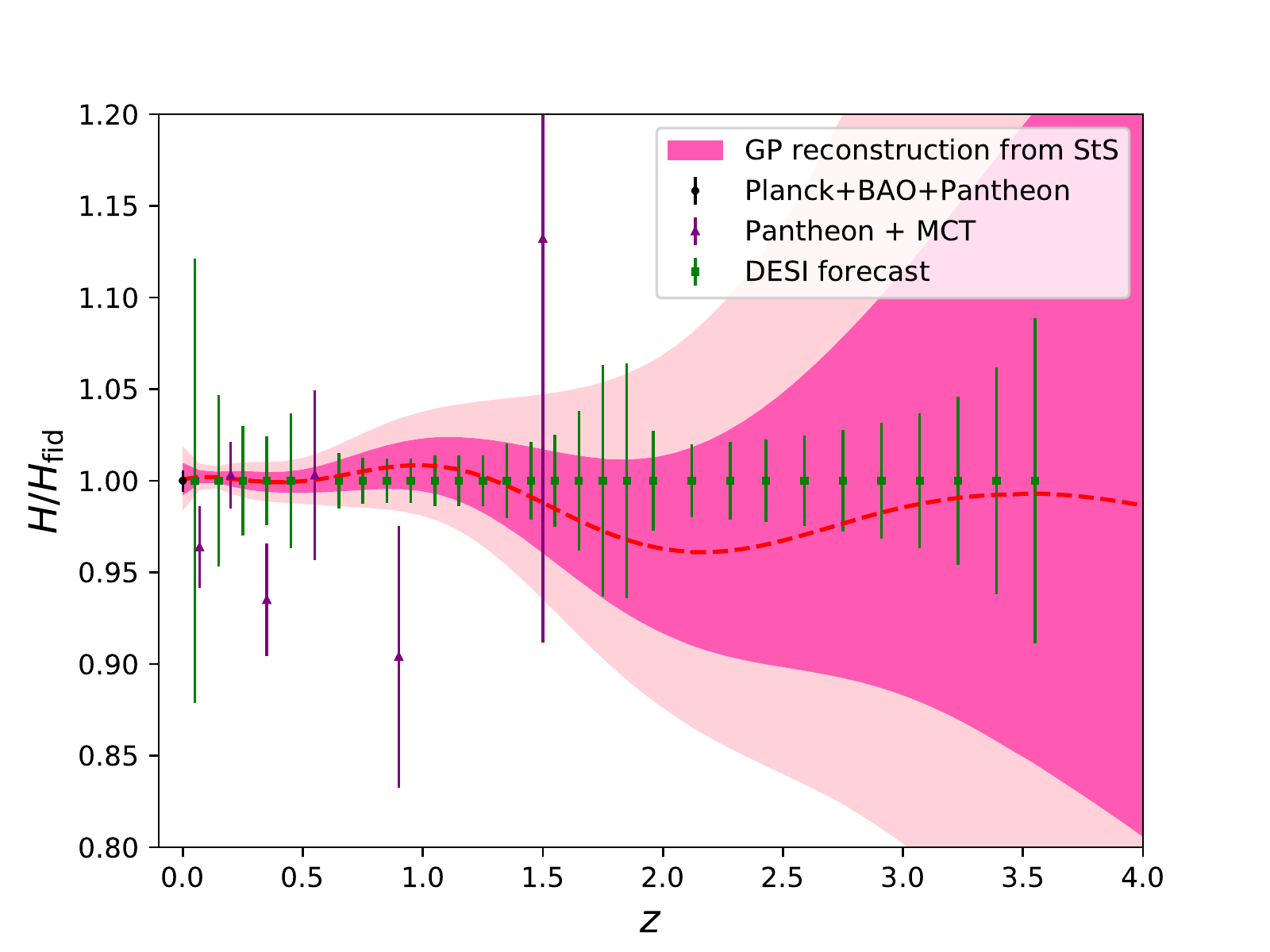}
\caption{The nonparametric GP reconstruction of Hubble parameter $H(z)$ from the combined mock standard sirens of future GW detector networks. Here we plot $H(z)$ relative to the fiducial $H_{\rm fid}(z)$ for a better illustration. The coloured areas show the regions which contain 68\% (pink) and $95\%$ (deep pink) of the probability. The dashed line is the mean of the reconstruction.  For comparison, we plot $H_0$ measurement from {\it Planck}+BAO+Pantheon, the measurements of $E(z)$ from Pantheon+MCT, and the forecast of the constraints of $H(z)$ from DESI. The error bars show $1\sigma$ errors. }
\label{fig:E_plot}
\end{figure}

\begin{figure}
\centering
\includegraphics[width=0.8\textwidth]{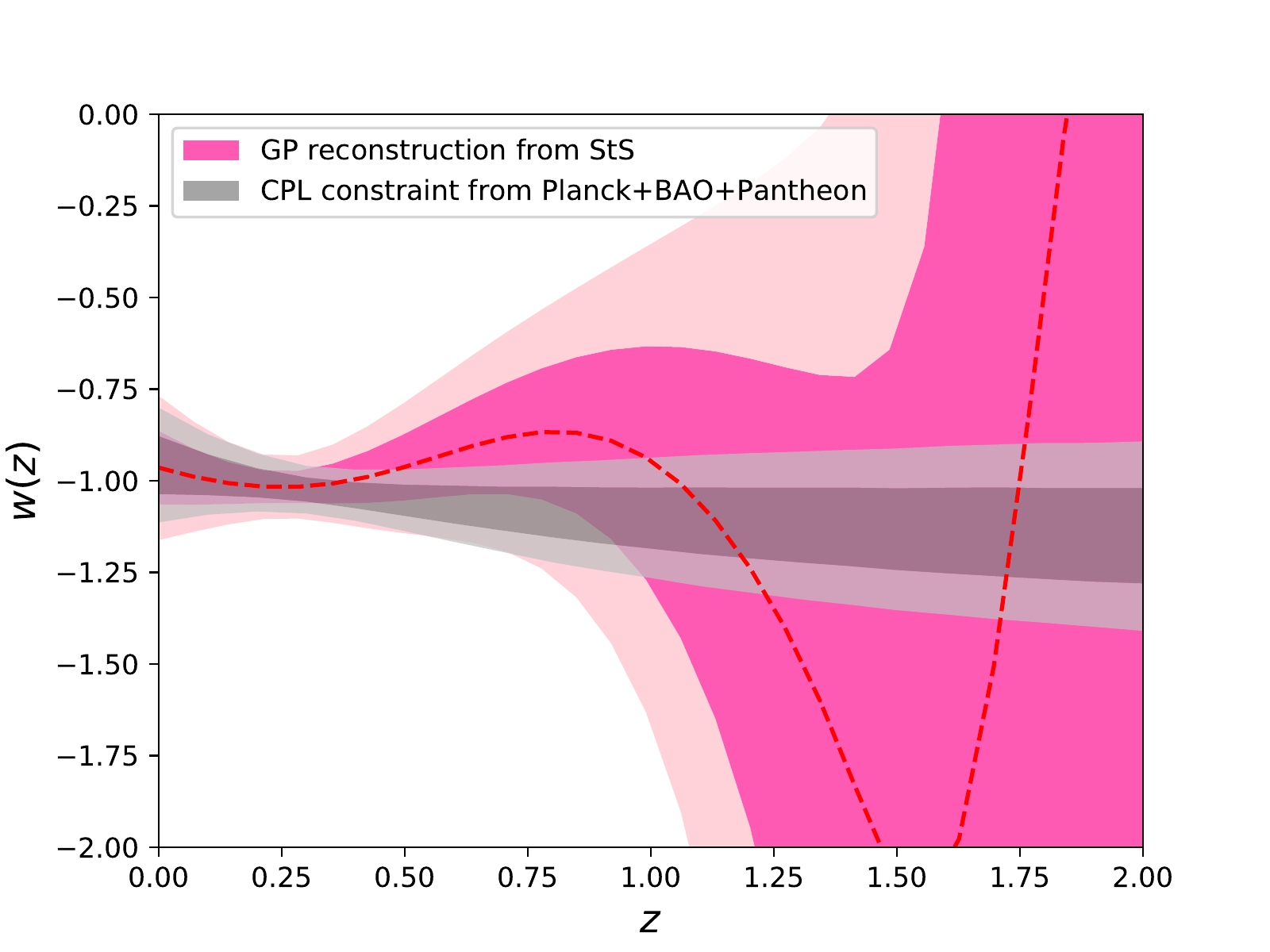}
\caption{The nonparametric GP reconstruction of the equation of state of dark energy $w(z)$ from the combined mock standard sirens of future GW detector networks. The coloured areas show the regions which contain 68\% and 95\% of the probability. The dashed line is the mean of the reconstruction.  For comparison, we plot $w(z)$ from the {\it Planck}+BAO+Pantheon MCMC posteriors of CPL model. }
\label{fig:w_plot}
\end{figure}

The GP reconstruction of Hubble parameter $H(z)$ from the standard sirens of future GW detector networks is shown in figure~\ref{fig:E_plot}. We plot $H(z)$ relative to the fiducial $H(z)_{\rm fid}$ for a better illustration.  As a comparison, we also plot the $H_0$ measurement from {\it Planck}+BAO+Pantheon, the six measurements of $E(z)$ from Pantheon+MCT~\cite{Riess:2017lxs}, the forecasted constraints of $H(z)$ achievable by DESI which covers 14,000 deg$^2$ in the future (see~\cite{Aghamousa:2016zmz} for details). We can see the nonparametric reconstruction of $H(z)$ from GW standard sirens alone can give a comparable measurement of Hubble constant with current {\it Planck}+BAO+Pantheon. For studying the expansion history of our Universe, the standard sirens alone can give a better constraints of $H(z)$ than Pantheon and future DESI at $z< 2$.
The reconstruction of the equation of state of dark energy $w(z)$ is shown in figure~\ref{fig:w_plot}. Again, as the comparison we plot the $w(z)$ from the {\it Planck}+BAO+Pantheon MCMC posteriors of CPL model. We can see with the standard sirens alone one can reconstruct the equation of state $w(z)$ in a nonparametric approach, which is comparable with the joint parametric constraints of {\it Planck}+BAO+Pantheon at redshift $z<0.25$.   
At present time we can only use ANN to reconstruct luminosity distance itself, in this paper we just show the GP results for the reconstructions of $H(z)$ and $w(z)$.

\begin{figure}
\centering
\includegraphics[width=0.49\textwidth]{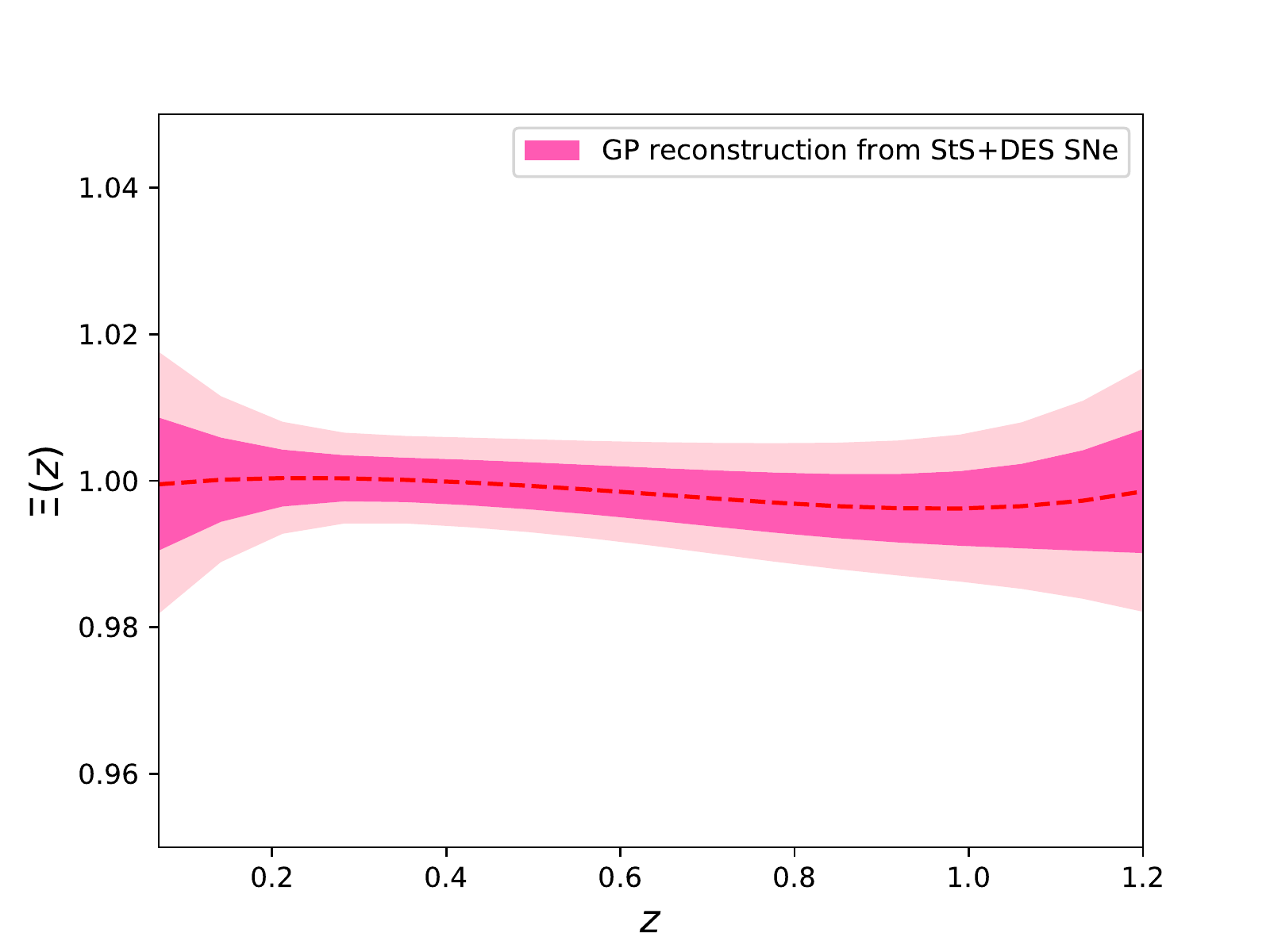}
\includegraphics[width=0.49\textwidth]{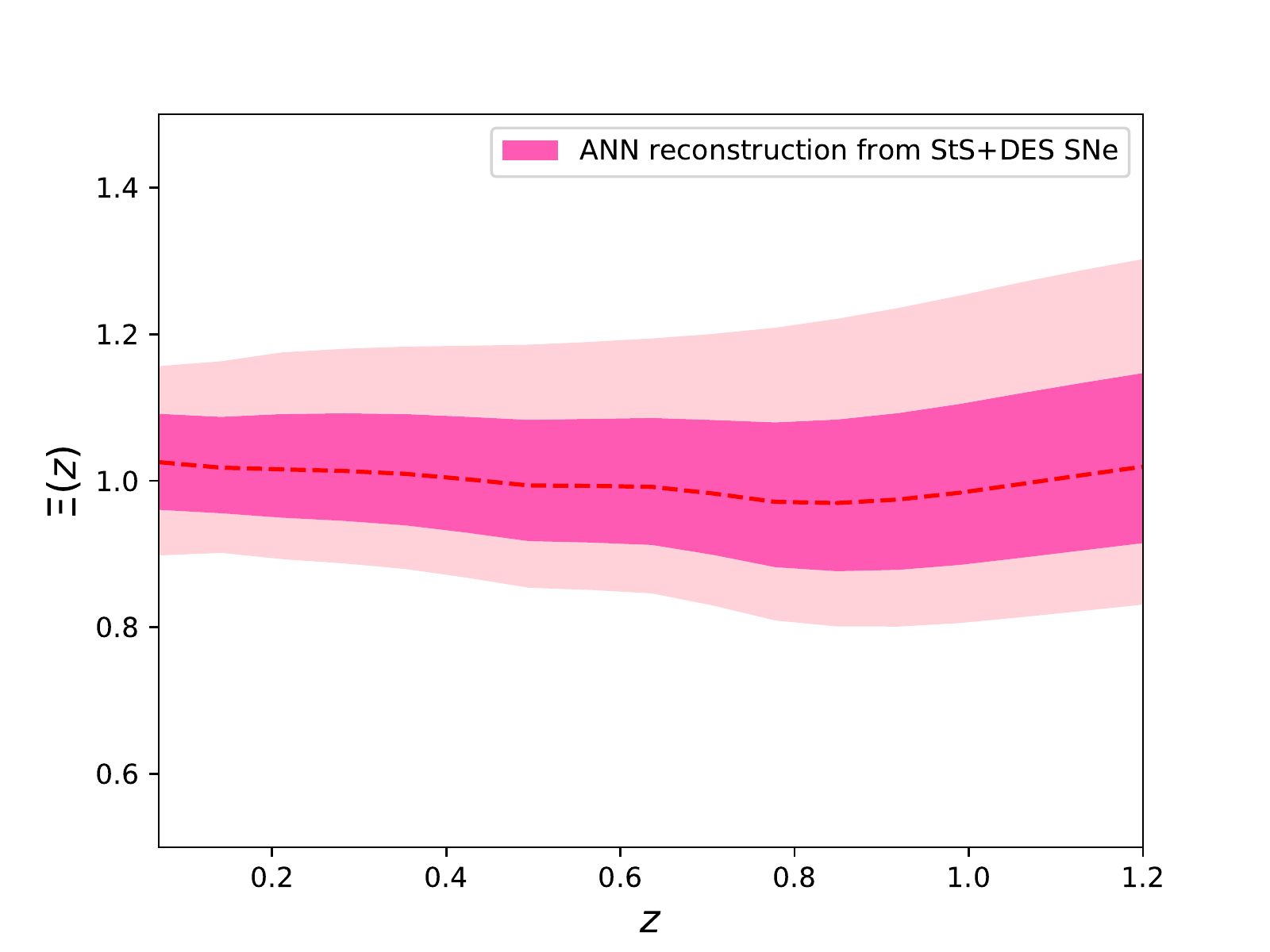}
\caption{The nonparametric GP (left) and ANN (right) reconstructions of $\Xi(z)$ from mock standard sirens and DES catalogues. The coloured areas show the regions which contain 68\% and 95\% of the probability. The dashed line is the mean of the reconstruction. }
\label{fig:Xi_plot}
\end{figure}

We then move to the constraint of MG with the modified GW propagation.  In section~\ref{sec:MGGW} we parameterize MG effect in GW propagation as~\eqref{eq:Xi}.  
Actually we can directly reconstruct $\Xi(z)$ from the reconstructions of $d_L^{\rm gw}(z)$ and $d_L^{\rm em}(z)$. Thus we can directly compare the reconstruction of $\Xi(z)$ with the predictions of the any explicit MG models in the redshift range covered by the data sets, which is a general and nonparametric method to test the gravity theory. $d_L^{\rm gw}(z)$ is reconstructed from the GW standard sirens as above. For $d_L^{\rm em}(z)$, we follow~\cite{Cai:2015zoa,Belgacem:2019zzu} and simulate SNe Ia data sets from the Dark Energy Survey (DES) strategy~\cite{Bernstein:2011zf}. Note when mocking up the DES SNe, we assume a well-determined Hubble constant  as the fiducial value. We adopt the same strategy as for the standard sirens to select the representative mock catalogue of DES SNe. We use GP and ANN to reconstruct $d_L(z)$ from GW standard siren and DES SNe separately. The kernel function of GP we choose is the Mat\'ern ($\nu=9/2$) as suggested in~\cite{Seikel:2013fda}. We adopt the optimal ANN model with 1 hidden layer and 4096 neurons which is derived in the optimizing of~\cite{Wang:2019vxv}. The ANN code we use is {\sc ReFANN}~\cite{Wang:2019vxv}. Since the largest redshift of DES SNe is 1.2, the final reconstruction of $\Xi(z)$ can be only up to $z=1.2$, as shown in figure~\ref{fig:Xi_plot}. 

\subsection{Summary of results}
Here we summarize our results of using standard sirens from future GW detector networks to study cosmology and modified gravity theory in section~\ref{sec:MCMC} and~\ref{sec:ML}.  We first use the Hubble diagram of standard sirens with MCMC to study the cosmological models such as $\Lambda$CDM, $w$CDM, CPL and the  MG theory through phenomenological parametrization of the GW propagation. The constraints of the parameters are summarized in table~\ref{tab:parameter}. In the base $\Lambda$CDM model, StS alone can constrain Hubble constant 
at the precision level of $0.34\%$, which is almost two times better than $0.6\%$ of current {\it Planck}+BAO+Pantheon. The joint {\it Planck}+BAO+Pantheon+StS can give $0.24\%$ precision for $H_0$. 
The errors of $\Omega_m$ given by StS, {\it Planck}+BAO+Pantheon and the joint are 2.74\%, 1.74\%, and 0.75\%, respectively.
If we extend the base model to introduce a constant equation of state of dark energy, using standard sirens alone one can constrain $w$ with an average error of 0.082, compared to 0.031 of {\it Planck}+BAO+Pantheon. The joint constraint would give an error of 0.019.  We can see from the contours of figure~\ref{fig:LCDM_MCMC} and~\ref{fig:wCDM_MCMC} that the significant shrink of the error for the joint constraint is due to the different directions of the parameter degeneracy between {\it Planck}+BAO+Pantheon and StS. For CPL model, the StS itself is not enough to give a tight constraint due to the large degeneracy between the parameters. However we can see StS can still give a better constraint of Hubble constant than {\it Planck}+BAO+Pantheon in CPL case. The inclusion of the StS to {\it Planck}+BAO+Pantheon can reduce the average errors of ($w_a$, $w_0$) from (0.08, 0.29) to (0.05, 0.18). 
For the phenomenological parametrization of GW propagation in MG, the combination of future GW StS and current {\it Planck}+BAO+Pantheon can constrain $w$ and $\Xi_0$ at the level of $1.9\%$ and $0.46\%$. 
For the extensions of the base $\Lambda$CDM model, using the GW standard sirens one can constrain the modified gravity through the propagation of GW much tighter than the dynamics of the dark energy. Thus, the sensitivity of GW detectors to the parameter $\Xi_0$ that characterizes modified GW propagation is better than to $w$, as has also been verified in~\cite{Belgacem:2018lbp,Belgacem:2019pkk,Belgacem:2019tbw}.

\begin{table}
\centering
\scalebox{0.7}{
\begin{tabular}{ll|ccccc} 
\hline\hline
\multicolumn{2}{l|}{\diagbox{Data and model}{Parameter}} & $H_0$           & $\Omega_m$                   & $w_0$                                          & $w_a$                                      & $\Xi_0$             \\ 
\hline
\multirow{3}{*}{StS}                     & $\Lambda$CDM  & $67.87\pm 0.23$ & $0.3069\pm 0.0084$           & --                                             & --                                         & --                  \\
                                         & $w$CDM        & $67.89\pm 0.50$ & $0.307^{+0.017}_{-0.015}$    & \multicolumn{1}{l}{$-1.005^{+0.088}_{-0.077}$} & --                                         & --                  \\
                                         & CPL   & $68.18\pm 0.60$ & $0.243^{+0.090}_{-0.036}$    & $-0.99\pm 0.12$                                & \multicolumn{1}{l}{$0.56^{+0.68}_{-0.24}$} & --                  \\ 
\hline
\multirow{3}{*}{Planck+BAO+Pantheon}     & $\Lambda$CDM  & $67.72\pm 0.40$ & $0.3104\pm 0.0054$           & --                                             & --                                         & --                  \\
                                         & $w$CDM        & $68.34\pm 0.81$ & $0.3057\pm 0.0075$           & $-1.028\pm 0.031$                              & --                                         & --                  \\
                                         & CPL   & $68.31\pm 0.82$ & $0.3065\pm 0.0077$           & $-0.957\pm 0.080$                              & $-0.29^{+0.32}_{-0.26}$                    & --                  \\ 
\hline
\multirow{4}{*}{Planck+BAO+Pantheon+StS} & $\Lambda$CDM  & $67.85\pm 0.16$ & $0.3086\pm 0.0023$           & --                                             & --                                         & --                  \\
                                         & $w$CDM        & $68.00\pm 0.25$ & $0.3084\pm 0.0023$           & $-1.015\pm 0.019$                              & --                                         & --                  \\
                                         & CPL   & $67.87\pm 0.39$ & $0.3101^{+0.0040}_{-0.0045}$ & $-0.992^{+0.051}_{-0.057}$                     & $-0.09^{+0.20}_{-0.16}$                    & --                  \\
                                         & MG            & $67.98\pm 0.31$ & $0.3087\pm 0.0033$           & $-1.015\pm 0.019$                              & --                                         & $0.9994\pm 0.0046$  \\ 
\hline\hline
StS                                      & GP            & $67.77\pm 0.60$             & --                           & $-0.964\pm 0.1$                                            & --                                         & --                  \\ 
\hline
\multirow{2}{*}{StS+DES SNe}             & GP            & --              & --                           & --                                             & --                                         & $0.998\pm 0.005$                 \\
                                         & ANN           & --              & --                           & --                                             & --                                         & $0.976^{+0.136}_{-0.113}$                 \\
\hline
\end{tabular}
}
\caption{The constraints of the parameters by different data combinations and approaches to study cosmology and modified gravity theory in this paper. ``StS'' and ``DES SNe'' are the mock standard sirens and DES supernovae. The numbers are the mean values with 68\% limits of the errors. Note for the nonparametric approach GP and ANN, we quote the constraints of $H_0$ and $w_0$ from the reconstructions of $H(z)$ and $w(z)$ at $z=0$. While $\Xi_0$ is derived from the reconstruction of $\Xi(z)$ where its error is the smallest.}
\label{tab:parameter}
\end{table}

In addition to the traditional model-fitting MCMC method, we also adopt the machine learning nonparametric reconstruction techniques like GP and ANN to reconstruct $d_L(z)$ from GW standard sirens. Using GP we reconstruct the Hubble parameter and equation of state as the function of redshift from StS alone. That provides a model-independent constraint of the expansion history of the Universe and the dynamics of dark energy back to an earlier time. From table~\ref{tab:parameter} we can see the errors of $H_0$ and $w_0$ from the nonparametric GP reconstructions are comparable with the specific model-fitting MCMC of {\it Planck}+BAO+Pantheon. Using the reconstruction of the EM luminosity distance from future DES SNe and GW luminosity distance from standard sirens, we reconstruct the $\Xi(z)$ as a function of redshift without assuming the parametric form. In this case, we use GP and ANN to reconstruct the luminosity distance from mock GW StS and DES SNe separately and then combine $d_L^{\rm gw}(z)$ and  $d_L^{\rm em}(z)$ together to derive the final reconstruction of $\Xi(z)$. From figure~\ref{fig:Xi_plot} we can see GP gives a much tighter reconstruction of $\Xi(z)$ than ANN. Table~\ref{tab:parameter} shows the constraint of $\Xi_0$ from GP reconstruction is comparable with MCMC. While using ANN the error of $\Xi_0$ is much larger.  The different tightness of the constraints from GP and ANN are duo to the basic logic and nature of these two techniques. The detailed investigation of the differences between GP and ANN is beyond the scope of this paper. Here we just would like to show an example of what we can obtain by applying GP and ANN as the nonparametric approach to the GW standard sirens.


\section{Discussions and prospects \label{sec:dis}}

The goal of this paper is to give a realistic construction of the Hubble diagram from the mock catalogues of standard sirens detected by future GW detector networks in the 2030s, and catch a glimpse of the its potential on studying cosmology and modified gravity theory. We focus on the GW events which are accompanied by the EM counterparts thus providing the information of the redshift of the host galaxies. These so called ``bright sirens'' are the more straightforward distance ladders than the statistical  ``dark sirens'' and without the calibration like SNe. Hence with standard sirens alone one can measure the Hubble constant. Our result shows the combined standard sirens for future GW detector networks, whose number is around order 200 in the 2030s,  can measure the Hubble constant 1.76 times better than current most precise EM experiments {\it Planck}+BAO+Pantheon in $\Lambda$CDM model. GW standard sirens would be one of the most promising tool to resolve the Hubble tension. On the other hand, with GW standard sirens measurements the modified gravity effect through the GW propagation is much more significant than introducing the dynamics of dark energy as extensions to the base $\Lambda$CDM, which makes the GW standard siren very powerful to test the gravity theory on the cosmic distance scales. Furthermore, the modified GW propagation is parameterized through a phenomenological form by introducing a function $\delta(z)$ in the friction term. This turns out to be a very general modification of GR, which also has the corresponding predictions for the specific MG theories. 
Thus one can implement a very general test of gravity theory without sticking to the specific MG theories.

Gaussian process has been maturely used in the nonparametric (cosmological model-independent) reconstruction of the cosmological parameters in the literature. In this paper we show GP has some advantages over the ANN in our study of cosmology and modified gravity theory. GP has more features like the reconstruction of the derivatives of the functions and the errors are much tighter than ANN. However, as a Bayesian reconstruction method GP interiorly assumes a kernel function (the correlation between each points) to smooth the function along the discrete points. The choice of the kernel function may influence the performance of the reconstruction (see e.g. some discussions in ~\cite{Busti:2014dua,Colgain:2021ngq} for using GP to determine $H_0$). In principle one needs to optimize the form of the kernel function from the mock data sets before the real applications. In this paper we directly adopt the optimal models of GP and ANN which have been suggested in the previous literature. We have showed what we can learn from the GW standard sirens on cosmology and modified gravity theory with the nonparametric approaches. This provides an intuitive and general test of the concordance cosmological model from local to high redshift. The fast-developing deep learning Neural Networks, which is however more comprehensive and powerful in the big data analysis than GP,  is now being widely used in cosmology and gravitational waves (data analysis and detections). We anticipate more research on these topics in the future.

When constructing the GW+EM catalogues, several assumptions and approximation have been made. In BNS case, we assumed the MD star formation rate with the exponential time delay distribution and the local BNS merger rate is predicted from the assumption of Gaussian distribution of NS mass. We did not take the spins of the binaries into account in this paper for simplification. This assumption would not influence too much on the estimation of the error of luminosity distance. The time dependence of the antenna response functions has not been included for ET/CE. We think it is a fair assumption considering the short time period of the observation in the BNS inspiral phase.
In LISA+Taiji case we only considered the light-seed model with delays (relative to the merger of host galaxy) included for the mergers of massive black holes, which are assumed to grow from the remnants of population III stars. While in the heavy-seed model the massive black holes are instead assumed to form from the collapse of protogalactic disks, with delays either included or not. In the literature people usually consider these three scenarios separately. From their works we can see the popIII model usually gives an average result among the three cases. In this paper, to construct a combined Hubble diagram from three GW detector networks in a conservative and realistic manner, we just adopt popIII model for MBHB. On the other hand, we assumed a very conservative improvement -- 10 times better for the localization of the GW when Taiji joining LISA. We also checked that if the average improvement is 100 times, then the number of MBHB standard sirens is around 80. Considering the fact that the total number of StS is around 200, we expect the gain 
of these data points would make little contribution to the overall performance. Taking account of the realistic synergetic observations of LISA and Taiji in the 2030s, however we think our estimation in this paper is reasonable. For a more specific investigation of the LISA-Taiji network, including the MBHB model, detector configuration, inference of waveform parameters, and the synergetic observation of the two detectors, we leave these for future research.

From table~\ref{tab:numbers} we can see the bright sirens are largely limited by the EM counterparts observations. In this paper the number of the bright sirens we estimated for the future GW detector networks in the 2030s could be over pessimistic or optimistic. We just performed a conservative estimation within current knowledge. However,  the error of the constraints of the cosmological parameters can be roughly estimated as $\propto1/\sqrt{N}$ with $N$ the number of the total StS events. On the other hand we can see the number of dark sirens is much more numerous than that of bright sirens, which shows the great potential of dark sirens on cosmology and modified gravity theory~\cite{Taylor:2011fs,DelPozzo:2015bna,Soares-Santos:2019irc,Mukherjee:2020hyn,Wang:2020dkc,Finke:2021aom}.

In the nonparametric approach to study the GW propagation in MG, one can also reconstruct $\delta(z)$ function which is in principle more fundamental than $\Xi(z)$~\cite{Belgacem:2019zzu}. However the reconstruction of EM luminosity distance is limited by the coverage range of SNe redshift. In this paper we do not reconstruct $\delta(z)$ from the combined Hubble diagram of StS due to the fact that the high redshift MBHB StS would not make contribution to the reconstruction at low redshift. To reconstruct either $\delta(z)$ or $\Xi(z)$ from the GW standard sirens at high redshift one needs the anchors of luminosity distance from EM candles in the same redshift range. Quasars (QSO) and gamma-ray bursts have been proposed as the standard candles to study cosmology, which can approach redshift $z\sim7$~\cite{Risaliti:2015zla,Demianski:2016zxi,Risaliti:2018reu,Wang:2013ha,Wang:2014fka,Cai:2018lye}. In particular, QSO is very promising, but it is not clear if it is a standard candle or can be standardised. Some analyses of the quasar data seem to give varying conclusions about the state of the standard cosmology~\cite{Melia:2019nev,Khadka:2019njj,Yang:2019vgk,Velten:2019vwo,Banerjee:2020bjq}. GW standard sirens are very suitable to check this~\cite{Speri:2020hwc}. We expect more research in the near future.


\acknowledgments

The author would like to thank Nicola Tamanini for helpful explanations and discussions on the mock catalogues of LISA standard sirens. We also thank Zong-Kuan Guo for sharing the sensitivity curve of Taiji program, Qing Yang for helpful discussions and Bin Hu for the comments on LISA-Taiji bright sirens. We thank Nicola Tamanini and Eoin \'O Colg\'ain for useful comments on the draft. This work is supported by an appointment to the YST Program at the APCTP through the Science and Technology Promotion Fund and Lottery Fund of the Korean Government, and the Korean Local Governments - Gyeongsangbuk-do Province and Pohang City.










\bibliographystyle{JHEP}
\bibliography{ref}

\end{document}